\documentclass[10pt,a4paper]{article}
\usepackage{graphicx}
\usepackage{color}
\usepackage{amsmath}
\usepackage{amssymb}
\usepackage{amscd}
\usepackage{bbm}
\usepackage{tikz}
\usepackage{framed}
\usepackage[normalem]{ulem}
\usetikzlibrary{positioning,arrows.meta,bending,calc}
\usepackage{hyperref}
\hypersetup{
    bookmarks=true,         
    colorlinks=true,       
    linkcolor=red,          
    citecolor=green,        
    filecolor=magenta,      
    urlcolor=cyan           
}
\usepackage[most]{tcolorbox}
\usepackage[utf8]{inputenc}
\usepackage{amsthm}
\usepackage{bbm} 
\usepackage{glossaries}
\usepackage[linesnumbered,lined,boxed,commentsnumbered]{algorithm2e}

\usepackage{marginnote}

\newtheorem{Theorem}{Theorem}

\newtheorem{Lemma}{Lemma}
\newtheorem{Proposition}{Proposition}
\newtheorem{Corollary}{Corollary}
\newtheorem{Remark}{Remark}

\newcounter{BigConst}                     

\newcounter{SmallConst}                     

\newcounter{gamma}                     

\newcounter{kappa}                     

\newcounter{delta}                     

\newcounter{theta}                     

\newcounter{L}                     

\newcounter{eps}                     

\let\savedamsthmendproof\endproof

\newcommand{\vertiii}[1]{{\bigl\vert\kern-0.25ex\bigl\vert\kern-0.25ex\bigl\vert #1 
    \bigr\vert\kern-0.25ex\bigr\vert\kern-0.25ex\bigr\vert}}

\newcommand{\vertiiii}[1]{{\bigl\vert\kern-0.25ex\bigl\vert\kern-0.25ex\bigl\vert\kern-0.25ex\bigl\vert #1 
    \bigr\vert\kern-0.25ex\bigr\vert\kern-0.25ex\bigr\vert\kern-0.25ex\bigr\vert}}

\def \endproof
{{\mbox{}\nolinebreak\hfill\rule{2mm}{2mm}\par\medbreak}}

\DeclareMathOperator*{\argmin}{argmin}
\DeclareMathOperator*{\supp}{supp}

\DeclareMathOperator{\Span}{span}

\DeclareMathOperator*{\KL}{KL}

\newcommand{\Proj}{{\mathrm{Proj}}}
\def\ds1{\textrm{1\kern-0.25emI}} 

\DeclareMathOperator{\Var}{\mathrm{Var}}

\newcommand \bE{{\mathbb E}}

\newcommand \bN{{\mathbb N}}

\newcommand \bP{{\mathbb P}}

\newcommand \bR{{\mathbb R}}

\newcommand{\bbeta}{{\boldsymbol{\beta}}}

\newcommand{\bu}{\boldsymbol{u}}
\newcommand{\bv}{\boldsymbol{v}}

\newcommand{\ve}{{\boldsymbol{e}}}

\newcommand{\vu}{{\boldsymbol{u}}}
\newcommand{\vv}{{\boldsymbol{v}}}

\newcommand{\vy}{{\boldsymbol{y}}}

\newcommand{\vxi}{{\boldsymbol{\xi}}}

\newcommand{\vzero}{\boldsymbol{0}}

\newcommand{\mysymbol}[3]{%
\newglossaryentry{#1}{%
      name={\ensuremath{#2}},%
      text={\ensuremath{#2}},%
      description={#3},%
      sort={#1}%
    }%
\expandafter\newcommand\expandafter{\csname smb#1\endcsname}{\gls{#1}}%
\expandafter\newcommand\expandafter{\csname #1\endcsname}{\ensuremath{#2}}%
}

\newcommand{\one}{\boldsymbol{1}}
\newcommand{\Cov}{\operatorname{Cov}}
\newcommand{\Bias}{\mathrm{Bias}}
\newcommand{\BIAS}{\mathrm{BIAS}}
\newcommand{\VAR}{\mathrm{VAR}}
\newcommand{\FSD}{\mathrm{FSD}}
\let\endproof\savedamsthmendproof
\usepackage[english]{babel}
\usepackage{enumitem}
\usepackage{microtype}
\usetikzlibrary{arrows.meta}
\usepackage{pgfplots}
\pgfplotsset{compat=1.18}
\numberwithin{equation}{section}
\begin{document}
\title{Instance-Specific Sharp Convergence Rate of the Add-$\lambda$ Estimator via Feature Space Decomposition}
\author{Zong Shang\thanks{School of Mathematics, Georgia Institute of Technology, Atlanta, GA, 30332-0160, USA. E-mail: \texttt{shangzong9@gmail.com}}}
\date{\today}
\maketitle
\begin{tikzpicture}[remember picture,overlay]
    \node[anchor=north] at ([xshift=-\hoffset,yshift=\dimexpr-\voffset-2pt\relax]current page.north) {\itshape Dedicated to the memory of my grandfather.};
\end{tikzpicture}
\begin{abstract}
    Using the Feature Space Decomposition framework, we establish non-asymptotic matching upper and lower bounds for the expected Kullback-Leibler estimation error of the Add-$\lambda$ estimator of a discrete probability distribution, for \textbf{any} signal, \textbf{any} dimension, \textbf{any} sample size, and \textbf{any} $\lambda>0$.
\end{abstract}
\setlength{\abovedisplayskip}{6pt plus 2pt minus 2pt}
\setlength{\belowdisplayskip}{6pt plus 2pt minus 2pt}
\setlength{\abovedisplayshortskip}{0pt plus 2pt}
\setlength{\belowdisplayshortskip}{4pt plus 2pt minus 2pt}


\section{Introduction}
This work examines a fundamental problem in mathematical statistics and information theory: the estimation of discrete probability distributions.

Let $N,p\in\bN_+$. Fix the canonical basis $\{\ve_1,\cdots,\ve_p\}$ of $\bR^p$. For any $\bbeta\in\bR^p$ and any $1\leq j\leq p$, we write $\beta_j=\langle\bbeta,\ve_j\rangle$. Let $\Delta_p=\{\bbeta\in[0,1]^p:\sum_{j=1}^p\beta_j=1\}$ be the probability simplex, and let $\bbeta^\star\in\Delta_p$ be a deterministic vector, called the signal. Let $X_1,\ldots,X_N$ be independent and identically distributed random variables with $\bP(X_i=j)=\beta_j^\star$, which form the training sample, and write
\begin{align*}
    N_j=\sum_{i=1}^N\mathbf1_{\{X_i=j\}},\quad \vy=\sum_{j=1}^p\frac{N_j}{N}\ve_j,\quad \mbox{ and }\quad \vxi=\vy-\bbeta^\star.
\end{align*}
The goal of this problem is to construct a discrete probability distribution $\widehat\bbeta$ from the observed training samples $\{X_i\}_{i=1}^N$ in order to estimate $\bbeta^\star$. The intuition is that for any $j\in[p]:=\{1,\cdots,p\}$, $\bE[N_j]=N\beta_j^\star$, so $y_j = \frac{N_j}{N}$ is an unbiased estimator of the $j$-th coordinate of $\bbeta^\star$. The quantity we use in this work to evaluate estimation performance is the Kullback-Leibler (KL) divergence, defined as follows. For nonnegative vectors $\vu,\vv\in\bR_{\geq 0}^p:=\{\bbeta\in\bR^p:\beta_j\geq 0,\forall 1\leq j\leq p\}$, set 
\begin{align}\label{eq:def_KL}
    \KL(\bu\|\bv)=\sum_{j=1}^p\Bigl(u_j\log\frac{u_j}{v_j}-u_j+v_j\Bigr),\qquad\mbox{ where }0\log\frac0v=0,\mbox{ and }u\log\frac u0=+\infty \mbox{ for }u>0.
\end{align}On $\Delta_p$, this agrees with the usual Kullback-Leibler divergence.

Estimating discrete measures is a standard unsupervised learning problem; see \cite[Section 29.4]{polyanskiywu}. Many estimators have been studied; see, for example, the following line of recent research \cite{han_minimax_2015,orlitsky_competitive_2015,hao_profile_2020,canonne_concentration_2023,han_besting_2025,han_optimal_2023,Bhattacharyya,hoeven_nearly_2026,mourtada2026}. This paper concerns the Add-$\lambda$ estimator, defined as follows. For a deterministic tuning parameter $\lambda\geq0$, consider the estimator
\begin{align}\label{eq:estimator}
    \begin{aligned}
        &\widehat{\bbeta}:=\widehat{\bbeta}^\lambda=\argmin_{\bbeta\in\Delta_p}\bigg\{-\frac1N\sum_{i=1}^N\log\beta_{X_i}-\frac\lambda N\sum_{j=1}^p\log\beta_j\bigg\},\quad \mbox{ that is }\quad  \widehat\beta_j=\frac{N_j+\lambda}{N+\lambda p},\quad 1\leq j\leq p.
    \end{aligned}
\end{align}
We call $\widehat\bbeta$ the \uline{Add-$\lambda$ estimator}. The choices $\lambda=0$, $\lambda=\frac{1}{2}$, and $\lambda=1$ give the maximum likelihood, Krichevsky-Trofimov, and Laplace estimators, respectively. In general, the parameter $\lambda$ may depend on $N,p$ and a prescribed confidence level. These estimators are widely used for estimating discrete probability distributions; see \cite{mourtada2026,hoeven_nearly_2026} for a detailed discussion and recent results.

In the literature, the asymptotic properties of $\widehat\bbeta$ have been thoroughly examined (\cite[Theorem 5.39]{vaartAsymptoticStatistics1998}), and oracle inequalities and minimax lower bounds for its estimation error $\KL(\bbeta^\star\|\widehat\bbeta)$ have also been established in \cite{canonne_concentration_2023,han_optimal_2023,Bhattacharyya,hoeven_nearly_2026,mourtada2026} and \cite[p. 664, Exercise VI.10]{polyanskiywu}. Most of these works adopt the minimax paradigm (\cite{waldStatisticalDecisionFunctions1949}), namely evaluating whether the statistical properties of $\widehat\bbeta$ attain information-theoretic optimality by examining whether the worst-case estimation error of an estimator over $\Delta_p$ matches, up to a constant, the worst-case estimation error over the model of the best estimator among all possible estimators. Another framework is the competitive framework, as in \cite{orlitsky_competitive_2015,han_besting_2025}, in which one hopes to establish a regret upper bound on the difference between the estimation error of $\hat\bbeta$ and the estimation error of a specially chosen class of competitive oracles on the given $\bbeta^\star$, so as to show that $\hat\bbeta$ is not too far from the competitive oracle. This paper studies a more refined question that complements the minimax paradigm, the competitive paradigm, and the existing literature:
\begin{framed}
    construct $r^2:=r^2(\bbeta^\star,p,N,\lambda)$ such that
    \begin{align*}
        \forall N,p\in\bN_+,\quad \forall \bbeta^\star\in\Delta_p,\quad\forall \lambda>0,\qquad \mathbb{E}[\KL(\bbeta^\star\|\widehat\bbeta)] \asymp r^2.
    \end{align*}
\end{framed}
Namely, we provide the \uline{sharp/instance-specific convergence rate} of the expected KL estimation error of the Add-$\lambda$ estimator for \textbf{any} signal $\bbeta^\star$, \textbf{any} dimension $p$, \textbf{any} sample size $N$, and \textbf{any} tuning parameter $\lambda$. We defer the new statistical insights into $\widehat\bbeta$ brought by the sharp convergence rate to Section~\ref{sec:comments}; we first introduce our notation.

\subsection{Notation}
This work uses the natural logarithm. We often use ``:='' to mean ``let ... be ...'', e.g., $[p]:=\{1,2,\cdots,p\}$ means letting $[p]$ be $\{1,2,\cdots,p\}$. We use $\one_{A}$ to denote the indicator of a set $A$, that is, $\one_A=1$ if $A$ occurs, and $0$ otherwise. For nonnegative real numbers $a,b$, we write $a\lesssim b$ if $a\leq cb$ for an absolute constant $c>0$, and write $a\gtrsim b$ for the reverse comparison. If both comparisons hold, we write $a\asymp b$. Subscripts indicate dependence on $\alpha,\iota,c,C$ whenever the comparison constants depend on these parameters.

\section{Main Result}

For any $\lambda>0$, define
\begin{align}\label{eq:decomposition}
    \begin{aligned}
        &J_\star(\lambda)=\bigg\{1\leq j\leq p:\beta_j^\star>\frac\lambda{4(N+\lambda p)}\bigg\},\qquad V_{J_\star}=\Span\{\ve_j:j\in J_\star\},\quad\mbox{ and }\quad V_{J_\star^c}=V_{J_\star}^\perp,
    \end{aligned}
\end{align}
When there is no ambiguity, we abbreviate $J_\star(\lambda)$ by $J_\star$.
For any $\bbeta\in\bR^p$, write $\bbeta_{J_\star}=\Proj_{V_{J_\star}}\bbeta$ for its orthogonal projection onto $V_{J_\star}$, and define $\bbeta_{J_\star^c}$ similarly. Let
\begin{align}\label{eq:fsd-components}
    \begin{aligned}&\mathrm{Bias}_{J_\star}=\sum_{j\in J_\star}\biggl[\beta_j^\star\log\frac{(N+\lambda p)\beta_j^\star}{N\beta_j^\star+\lambda}+\frac{N\beta_j^\star+\lambda}{N+\lambda p}-\beta_j^\star\biggr],\quad \mathrm{Bias}_{J_\star^c}=\frac{\lambda|J_\star^c|}{N+\lambda p},\quad\mbox{ and }\\ &\\ &\mathrm{VAR}^{\mathrm{FSD}}=\sum_{j\in J_\star}\biggl[\frac{N(\beta_j^\star)^2(1-\beta_j^\star)}{(N\beta_j^\star+\lambda)(N\beta_j^\star+\lambda+1)}+\mathbf1_{\{\lambda<1\}}\beta_j^\star(1-\beta_j^\star)^N\log\biggl(1+\frac{N\beta_j^\star}{\lambda}\biggr)\biggr].\end{aligned}
\end{align}
We defer the interpretation of these three quantities to Section~\ref{sec:geometric_perspective}. We define
\begin{align}\label{eq:rate}
    \mathrm{BIAS}^{\mathrm{FSD}}=\mathrm{Bias}_{J_\star}+\mathrm{Bias}_{J_\star^c},\quad\mbox{ and }\quad r^2(V_{J_\star},V_{J_\star^c})=\mathrm{VAR}^{\mathrm{FSD}}+\mathrm{BIAS}^{\mathrm{FSD}}.
\end{align}
The following is the main result of this work, whose proof may be found in Section~\ref{sec:proof_main}.
\begin{Theorem}
    \label{thm:main}
    For every $N,p\in\bN_+$, $\bbeta^\star\in\Delta_p$ and deterministic $\lambda>0$,
    \begin{align}
        \begin{aligned}
            &\frac1{14}r^2(V_{J_\star},V_{J_\star^c})\leq\bE\KL(\bbeta^\star\|\widehat{\bbeta})\leq12r^2(V_{J_\star},V_{J_\star^c}).\label{eq:main-expectation}
        \end{aligned}
    \end{align}
\end{Theorem}
The method used in this work to prove Theorem~\ref{thm:main} is the Feature Space Decomposition (FSD) method developed by G. Lecué and the author. We refer interested readers to the author's PhD thesis \cite{shangFeatureSpaceDecomposition2026} for a more comprehensive understanding of this framework and method. In terms of proof methodology, even though $\widehat\bbeta$ can be written as a regularized empirical risk minimizer (RERM), this paper does not, as in previous works in the FSD line \cite{lecueGeometricalViewpointBenign2024,gavrilopoulosGeometricalAnalysisKernel2025,adamczakFeatureSpaceDecompositioninpreperation,lecuesharp2026}, characterize its estimation error by exploring its effective regularization. This is because each coordinate of the Add-$\lambda$ estimator is separable, namely $\widehat\beta_j = \frac{N_j+\lambda}{N+\lambda p}$, so that even if $\widehat\bbeta_{J_\star}$ can be written as an RERM, $V_{J_\star^c}$ has essentially no effect on $\widehat\bbeta_{J_\star}$. Instead, this paper analyzes each coordinate of the Add-$\lambda$ estimator directly in order to obtain the sharp convergence rate.

\begin{Remark}
    Since $\widehat\bbeta$ is an affine function of $\vy$, we can define the bias and variance of its KL estimation error in the classical way as $\BIAS^{\mathrm{cls}}:=\KL(\bbeta^\star\|\mathbb{E}[\widehat\bbeta])$ and $\VAR^{\mathrm{cls}}:=\sum_{j=1}^p\beta_j^\star(\log(N\beta_j^\star+\lambda) - \mathbb{E}[\log(N_j+\lambda)])$, respectively. Then by $\bE[\widehat\beta_j] = \frac{N\beta_j^\star+\lambda}{N+\lambda p}$,
    \begin{align*}
        \mathbb{E}[\KL(\bbeta^\star\|\widehat\bbeta)] &=\sum_{j=1}^p\beta_j^\star\Bigl( \log\beta_j^\star - \bE[\log\widehat\beta_j] \Bigr) = \sum_{j=1}^p\beta_j^\star\log\frac{\beta_j^\star}{\bE[\widehat\beta_j]} + \sum_{j=1}^p\beta_j^\star\Bigl( \log\bE[\widehat\beta_j] - \bE[\log\widehat\beta_j] \Bigr) \\
        &= \BIAS^{\mathrm{cls}} + \VAR^{\mathrm{cls}}.
    \end{align*}
    One can check that for any $N,p\in\mathbb{N}_+$, any $\bbeta^\star\in\Delta_p$, and any $\lambda>0$, we have $\BIAS^{\mathrm{FSD}}\asymp \BIAS^{\mathrm{cls}}$; and $\VAR^{\mathrm{FSD}} \lesssim \VAR^{\mathrm{cls}} \lesssim \VAR^{\mathrm{FSD}} + \Bias_{J_\star^c}$, and hence $\BIAS^{\mathrm{cls}} + \VAR^{\mathrm{cls}}\asymp r^2(V_{J_\star},V_{J_\star^c})$. Thus, we account for part of the classical variance through the bias term.
\end{Remark}

\section{Comments}\label{sec:comments}
In this section, we comment on Theorem~\ref{thm:main}. The comments concern the following aspects:
\begin{enumerate}
    \item a geometric perspective on the estimation error of the Add-$\lambda$ estimator;
    \item the novel perspective beyond the minimax paradigm provided by Theorem~\ref{thm:main}.
\end{enumerate}

\subsection{A geometric perspective on the estimation error of the Add-\texorpdfstring{$\lambda$}{λ} estimator}\label{sec:geometric_perspective}

In this subsection, we provide a geometric perspective on the estimation error of the Add-$\lambda$ estimator through the FSD framework. 

The FSD framework studies the non-asymptotic properties of the estimation error and the population excess risk. We refer the interested reader to the author's PhD thesis \cite{shangFeatureSpaceDecomposition2026} for a comprehensive overview. It states that, \textit{ fundamentally, there are exactly two different reasons for the estimation error not to be too large:
\begin{enumerate}
    \item the cancellation between the signal and the estimator on $V_{J_\star}$, and 
    \item the smallness of the signal and the estimator on $V_{J_\star^c}$. 
\end{enumerate}}
The classical estimation error examined by the FSD framework is the $L^2$ estimation error, namely the (squared) $L^2$ distance $\|\widehat f - f^\star\|_{L^2}^2$ between an estimator $\widehat f$ and a signal $f^\star\in L^2$. Since $L^2$ is a Hilbert space, these two fundamental reasons can be understood, respectively, as:
\begin{enumerate}
    \item the angle between the projection of $\widehat f$ on $V_{J_\star}$ and that of $f^\star$ being small; 
    \item the sum of the squared $L^2$ norms of the projection of $\widehat f$ on $V_{J_\star^c}$ and that of $f^\star$ being small; see, e.g., \cite[Section 1.5]{shangFeatureSpaceDecomposition2026}. 
\end{enumerate}
Hence, the contribution of $V_{J_\star}$ to the $L^2$ estimation error not being too large is that the projection of $\widehat f$ onto $V_{J_\star}$ estimates the projection of $f^\star$ onto $V_{J_\star}$, and thus $V_{J_\star}$ is called the \uline{estimation subspace}; whereas the contribution of $V_{J_\star^c}$ to the $L^2$ estimation error not being too large \textit{does not} come from estimation (since the angle between the two does not necessarily shrink), but from the smallness of the sum of the squared $L^2$ norms of the two, and hence we tentatively call this subspace the \uline{free subspace}.

By clarifying the two fundamentally different contributions to the estimation error, the FSD framework provides redefinitions of the most basic concepts of mathematical statistics, including consistency (\cite{adamczakFeatureSpaceDecompositioninpreperation}) and the bias-variance decomposition (\cite{lecuesharp2026}), as well as a general method for establishing the sharp convergence rate of the estimation error/population excess risk. 

The estimation error considered in this work is defined through the KL divergence, and hence does not itself provide an inner product with which to define angles, as the $L^2$ structure does. Taking the KL divergence estimation error studied in this work as an example, we show that
\begin{framed}
    the two fundamentally different mechanisms identified by the FSD framework for controlling the estimation error/population excess risk also apply to \textbf{estimation errors other than the $L^2$ estimation error}, and for \textbf{unsupervised learning problems}.
\end{framed}
To this end, we introduce a mapping that maps $\Delta_p$ into a space in which angles and lengths can be conveniently defined, that is, we use the following mapping to replace the estimation problem in $\Delta_p$ by an estimation problem in $S_2^{p-1}(\vzero;2):=\{\vv\in\mathbb{R}^p:\,\|\vv\|_2=2\}$, where $\|\cdot\|_2$ is the Euclidean norm. Let
\begin{align*}
    \Delta_p\ni \bbeta\longmapsto 2\sqrt{\bbeta} := (2\sqrt{\beta_j})_{j=1}^p \in S_2^{p-1}(\vzero;2)\cap \bR_{\geq 0}^p.
\end{align*}This mapping maps each discrete probability distribution in $\Delta_p$ to the part of the Euclidean sphere of radius $2$ in the nonnegative orthant\footnote{This mapping is in fact the classical square-root embedding in information geometry, see \cite{amari_information_2016}; however we do not make use of information geometry in this work except this mapping.}. This returns us to a space in which angles can be visualized; see Figure~\ref{fig:geometric_perspective}. 
In addition, on $S_2^{p-1}(\vzero;2)\cap\mathbb{R}_{\geq0}^p$, we can also relate the angle between two vectors $2\sqrt{\bbeta}$ and $2\sqrt{\bbeta'}$ to the KL divergence, namely
\begin{align*}
    \forall \bbeta'\in\Delta_p\text{ with }\beta_j'>0\ (1\leq j\leq p),\qquad \lim_{\bbeta\to\bbeta'}\frac{\KL(\bbeta\|\bbeta')}{\bigl(\angle(2\sqrt{\bbeta},2\sqrt{\bbeta'})\bigr)^2}=2,
\end{align*}see Lemma~\ref{lem:angle} for the proof.
Below we show how this provides us with a geometric perspective for understanding the estimation error of the Add-$\lambda$ estimator.
In Section~\ref{sec:proof_main}, we prove the identity
\begin{align}\tag{\ref{eq:decomposition_KL}}
    \KL(\bbeta^\star\|\widehat\bbeta) = \KL(\bbeta_{J_\star}^\star\|\widehat\bbeta_{J_\star}) + \KL(\bbeta_{J_\star^c}^\star\|\widehat\bbeta_{J_\star^c}).
\end{align}
\begin{enumerate}
    \item We prove in Lemma~\ref{lem:estimation-upper} that
    \begin{align*}
        \bE[\KL(\bbeta_{J_\star}^\star\|\widehat\bbeta_{J_\star})]\asymp \Bias_{J_\star} + \VAR^{\mathrm{FSD}}.
    \end{align*}Therefore, on the estimation subspace $V_{J_\star}$, the coordinate projection $\bbeta_{J_\star}^\star$ of $\bbeta^\star$ is estimated, in the KL sense, by the coordinate projection $\widehat\bbeta_{J_\star}$ of $\widehat\bbeta$. Here, since $\|2\sqrt{\bbeta_{J_\star}^\star}\|_2, \|2\sqrt{\widehat\bbeta_{J_\star}}\|_2 \le 2$, to focus on the angle, we take the intersection points on $S_2^{p-1}(\vzero;2)\cap \mathbb{R}_{\geq0}^p$ of the extensions of the lines from $\vzero$ to $2\sqrt{\widehat\bbeta_{J_\star}}$ and to $2\sqrt{\bbeta_{J_\star}^\star}$ as $\bbeta$ and $\bbeta'$, namely $\bbeta:=\frac{2\sqrt{\widehat\bbeta_{J_\star}}}{\sqrt{\sum_{j\in J_\star}\widehat\beta_j}}$ and $\bbeta':=\frac{2\sqrt{\bbeta_{J_\star}^\star}}{\sqrt{\sum_{j\in J_\star}\beta_j^\star}}$. Hence we have
    \begin{align*}
        \angle\biggl( \frac{2\sqrt{\widehat\bbeta_{J_\star}}}{\sqrt{\sum_{j\in J_\star}\widehat\beta_j}}, \frac{2\sqrt{\bbeta_{J_\star}^\star}}{\sqrt{\sum_{j\in J_\star}\beta_j^\star}} \biggr) = \angle(2\sqrt{\widehat\bbeta_{J_\star}},2\sqrt{\bbeta_{J_\star}^\star})  = \Theta(\sqrt{\KL(\bbeta_{J_\star}^\star\|\widehat\bbeta_{J_\star})}).
    \end{align*}That is to say, on $V_{J_\star}$, the reason why the KL estimation error is not too large is still the cancellation between $\widehat\bbeta_{J_\star}$ and $\bbeta_{J_\star}^\star$, manifested as a small angle between them, see Figure~\ref{fig:geometric_perspective}.

    \item We prove in Lemma~\ref{lem:free-upper} that
    \begin{align*}
        \KL(\bbeta_{J_\star^c}^\star\|\widehat\bbeta_{J_\star^c}) \asymp \|2\sqrt{\widehat\bbeta_{J_\star^c}}\|_2^2,\mbox{ a.s.,} \quad\mbox{ and }\quad \bE[\KL(\bbeta_{J_\star^c}^\star\|\widehat\bbeta_{J_\star^c})]\asymp \Bias_{J_\star^c}.
    \end{align*}On the free subspace, the KL estimation error is described by the bias term $\Bias_{J_\star^c}$. The mass $\frac{\lambda|J_\star^c|}{N+\lambda p}$ introduced by smoothing is equivalent to the expected squared Euclidean norm of $2\sqrt{\widehat\bbeta_{J_\star^c}}$, and thus describes its contribution to the KL estimation error. We emphasize that the equivalence $\KL(\bbeta_{J_\star^c}^\star\|\widehat\bbeta_{J_\star^c}) \asymp \|2\sqrt{\widehat\bbeta_{J_\star^c}}\|_2^2$ may initially seem counterintuitive, since its right-hand side does not explicitly contain $\bbeta_{J_\star^c}^\star$. It follows directly from the definition of $J_\star^c$. This is because once $\bbeta^\star\in\Delta_p$ and $\lambda>0$ are fixed, the definition determines $J_\star^c$ such that for every $j \in J_\star^c$, we have $\widehat\beta_j \geq 4\beta_j^\star$ (see \eqref{eq:separation}). This can be regarded as a second explanation for why no estimation occurs on $V_{J_\star^c}$ — for every $j \in J_\star^c$, there is a non-trivial gap between $\widehat\beta_j$ and $\beta_j^\star$. As a consequence,
    \begin{align}\tag{\ref{eq:each_coordinate}}
        \KL(\bbeta_{J_\star^c}^\star\|\widehat\bbeta_{J_\star^c}) = \sum_{j\in J_\star^c}\widehat\beta_j\Bigl( 1 - \frac{\beta_j^\star}{\widehat\beta_j} + \frac{\beta_j^\star}{\widehat\beta_j}\log\frac{\beta_j^\star}{\widehat\beta_j} \Bigr) \asymp  \|2\sqrt{\widehat\bbeta_{J_\star^c}}\|_2^2.
    \end{align}
    That is to say, on $V_{J_\star^c}$, the reason why the KL estimation error is not too large is the smallness of $\widehat\bbeta_{J_\star^c}$, manifested as the squared Euclidean norm of $2\sqrt{\widehat\bbeta_{J_\star^c}}$ being small; while the squared Euclidean norm of $2\sqrt{\bbeta_{J_\star}^\star}$ has no effect on the estimation error, see Figure~\ref{fig:geometric_perspective}. 
\end{enumerate}

\begin{figure}[htbp]
\centering
\definecolor{fsdred}{RGB}{220,22,35}
\definecolor{fsdblue}{RGB}{0,85,225}
\definecolor{fsdpink}{RGB}{211,45,151}
\begin{tikzpicture}[x=1cm,y=1cm,>=Latex,line cap=round,line join=round]
\coordinate (O) at (0,0);

\pgfmathsetmacro{\uxh}{1.32}
\pgfmathsetmacro{\uyh}{0.51}
\pgfmathsetmacro{\uzh}{0.35}
\pgfmathsetmacro{\urh}{sqrt(\uxh*\uxh+\uyh*\uyh)}
\pgfmathsetmacro{\uxt}{1.54}
\pgfmathsetmacro{\uyt}{0.80}
\pgfmathsetmacro{\uzt}{sqrt(4-\uxt*\uxt-\uyt*\uyt)}
\pgfmathsetmacro{\urt}{sqrt(\uxt*\uxt+\uyt*\uyt)}

\coordinate (Phat) at ({2.10*\uxh-\uyh},{-0.15*\uxh-1.25*\uyh});
\coordinate (Ptrue) at ({2.10*\uxt-\uyt},{-0.15*\uxt-1.25*\uyt});
\coordinate (Fhat) at (0,{2*\uzh});
\coordinate (Bhat) at ({2.10*\uxh-\uyh},{-0.15*\uxh-1.25*\uyh+2*\uzh});
\coordinate (Btrue) at ({2.10*\uxt-\uyt},{-0.15*\uxt-1.25*\uyt+2*\uzt});
\coordinate (Qhat) at ({(4.20*\uxh-2.00*\uyh)/\urh},{(-0.30*\uxh-2.50*\uyh)/\urh});
\coordinate (Qtrue) at ({(4.20*\uxt-2.00*\uyt)/\urt},{(-0.30*\uxt-2.50*\uyt)/\urt});

\fill[fsdred,opacity=0.12] (-2.20,0.50)--(5.25,-0.50)--(3.00,-3.20)--(-4.45,-2.20)--cycle;
\draw[fsdred!72,line width=0.63pt] (-2.20,0.50)--(5.25,-0.50)--(3.00,-3.20)--(-4.45,-2.20)--cycle;
\node[fsdred,anchor=west,font=\large] at (3.47,-3.02) {$V_{J_\star}$};

\draw[fsdblue,line width=1.0pt,-{Latex[length=2.7mm]}] (O)--(0,4.75);
\node[fsdblue,anchor=west,font=\large] at (0.12,4.57) {$V_{J_\star^c}$};
\draw[fsdblue,line width=1.35pt,-{Latex[length=2.5mm]}] (O)--(Fhat);
\fill[fsdblue] (Fhat) circle[radius=1.7pt];
\node[fsdblue,anchor=east,font=\footnotesize] at (-0.16,0.82) {$2\sqrt{\widehat{\boldsymbol\beta}_{J_\star^c}}$};

\draw[fsdblue,densely dashed,line width=0.5pt] (Phat)--(Bhat);
\draw[fsdred,densely dashed,line width=0.5pt] (Fhat)--(Bhat);
\draw[black!55,densely dashed,line width=0.5pt] (Ptrue)--(Btrue);

\draw[black,line width=1.12pt,-{Latex[length=2.5mm]}] (O)--(Bhat);
\draw[black,line width=1.12pt,-{Latex[length=2.5mm]}] (O)--(Btrue);
\fill[black] (Bhat) circle[radius=1.7pt];
\fill[black] (Btrue) circle[radius=1.7pt];
\node[black,anchor=west,font=\footnotesize] at (2.45,-0.05) {$2\sqrt{\widehat{\boldsymbol\beta}}$};
\node[black,anchor=west,font=\footnotesize] at (0.83,0.63) {$2\sqrt{\boldsymbol\beta^\star}$};

\draw[fsdred,line width=1.2pt,-{Latex[length=2.3mm]}] (O)--(Phat);
\draw[fsdred,line width=1.2pt,-{Latex[length=2.3mm]}] (O)--(Ptrue);
\fill[fsdred] (Phat) circle[radius=1.7pt];
\fill[fsdred] (Ptrue) circle[radius=1.7pt];
\node[fsdred,anchor=west,font=\footnotesize] at (2.40,-0.65) {$2\sqrt{\widehat{\boldsymbol\beta}_{J_\star}}$};
\node[fsdred,anchor=north,font=\footnotesize] at (1.18,-1.02) {$2\sqrt{\boldsymbol\beta_{J_\star}^\star}$};

\draw[fsdred,line width=0.85pt] (O) ++(-26.83:0.57) arc[start angle=-26.83,end angle=-20.27,radius=0.57];

\draw[fsdpink,densely dashed,line width=0.65pt] (Phat)--(Qhat);
\draw[fsdpink,densely dashed,line width=0.65pt] (Ptrue)--(Qtrue);
\draw[fsdpink,line width=0.5pt] (Qhat)--(4.30,-1.18)--(5.30,-1.18);
\node[fsdpink,anchor=west,font=\footnotesize] at (5.40,-1.18) {$\displaystyle\frac{2\sqrt{\widehat{\boldsymbol\beta}_{J_\star}}}{\sqrt{\sum_{j\in J_\star}\widehat\beta_j}}$};
\draw[fsdpink,line width=0.5pt] (Qtrue)--(2.64,-3.50)--(0.95,-3.50);
\node[fsdpink,anchor=north,font=\footnotesize] at (1.10,-3.66) {$\displaystyle\frac{2\sqrt{\boldsymbol\beta_{J_\star}^\star}}{\sqrt{\sum_{j\in J_\star}\beta_j^\star}}$};

\draw[black,line width=1.48pt] plot[domain=0:90,samples=100,variable=\t] ({4.20*cos(\t)-2.00*sin(\t)},{-0.30*cos(\t)-2.50*sin(\t)});
\draw[black,line width=1.48pt] plot[domain=0:90,samples=100,variable=\t] ({4.20*cos(\t)},{-0.30*cos(\t)+4.00*sin(\t)});
\draw[black,line width=1.48pt] plot[domain=0:90,samples=100,variable=\t] ({-2.00*cos(\t)},{-2.50*cos(\t)+4.00*sin(\t)});

\fill[fsdpink] (Qhat) circle[radius=2.7pt];
\fill[fsdpink] (Qtrue) circle[radius=2.7pt];

\node[black,anchor=west,font=\footnotesize] at (4.32,1.08) {$S_2^{p-1}(\boldsymbol 0;2)$};
\fill[black] (O) circle[radius=1.6pt];
\end{tikzpicture}
\caption{After mapping $\Delta_p$ to $S_2^{p-1}(\vzero;2)\cap \mathbb{R}_{\geq0}^p$ via the map $\bbeta\mapsto 2\sqrt{\bbeta}$, we perform a feature space decomposition of the feature space $\mathbb{R}^p$, namely $\mathbb{R}^p = {\color{fsdred}V_{J_\star}}\oplus^\perp {\color{fsdblue}V_{J_\star^c}}$. Here, we use {\color{fsdred}red} to denote ${\color{fsdred}V_{J_\star}}$ and {\color{fsdblue}blue} to denote ${\color{fsdblue}V_{J_\star^c}}$. Correspondingly, $2\sqrt{\widehat\bbeta}$ can also be orthogonally decomposed onto these two subspaces, and we use {\color{fsdred}red} and {\color{fsdblue}blue} arrows to denote ${\color{fsdred}2\sqrt{\widehat\bbeta_{J_\star}}}$ and ${\color{fsdblue}2\sqrt{\widehat\bbeta_{J_\star^c}}}$, respectively. For the (mapped) signal $2\sqrt{\bbeta^\star}$, we draw only its projection ${\color{fsdred}2\sqrt{\bbeta_{J_\star}^\star}}$ onto the estimation subspace ${\color{fsdred}V_{J_\star}}$. The FSD framework states that $\widehat\bbeta$ has a small estimation error in the KL divergence sense with respect to $\bbeta^\star$ if and only if, on the estimation subspace, the angle between the two projected vectors is small (cancellation between ${\color{fsdred}2\sqrt{\widehat\bbeta_{J_\star}}}$ and ${\color{fsdred}2\sqrt{\bbeta_{J_\star}^\star}}$), and on the free subspace, the Euclidean norm of ${\color{fsdblue}2\sqrt{\widehat\bbeta_{J_\star^c}}}$ is small (smallness of $\|{\color{fsdblue}2\sqrt{\widehat\bbeta_{J_\star^c}}}\|_2$).}
\label{fig:geometric_perspective}
\end{figure}

\subsection{Beyond minimax optimality}\label{subsec:beyond-minimax}

In this subsection, we attempt to answer a question that may arise at the outset: \textit{why do we need to characterize the convergence rate of $\KL(\bbeta^\star\|\widehat\bbeta)$ itself, rather than establishing its minimax lower bound and an oracle inequality as an upper bound?} 

The minimax optimality paradigm, to the best of our knowledge, was first proposed by A. Wald (\cite{waldStatisticalDecisionFunctions1949}) under the influence of game theory. Its motivation is that the signal the statistician wishes to estimate is often chosen by nature, and the statistician knows nothing about nature's choice, and hence must play a game against nature. This requires that, even for the worst signal, the estimator designed by the statistician still exhibit the best possible statistical properties (i.e., a small estimation error). Here, ``best possible'' means that the worst-case estimation error of this estimator should be equivalent, up to a multiplicative constant, to the worst-case estimation error obtained by taking the infimum over the class of all Borel measurable functions.

Specifically for the discrete probability distribution estimation problem considered in this work, the minimax paradigm evaluates an estimator $\tilde \bbeta$ by the following criterion: whether it satisfies an oracle inequality whose residual term coincides with the minimax lower bound over $\Delta_p$, namely $$\sup_{\bbeta^\star\in\Delta_p}\mathbb{E}_{\bbeta^\star}\KL(\bbeta^\star\|\tilde\bbeta) \lesssim \inf_{\bar\bbeta}\sup_{\bbeta^\star\in\Delta_p}( \mathbb{E}_{\bbeta^\star}\KL(\bbeta^\star\|\bar\bbeta) ),$$ where the infimum is taken over all measurable $\bar\bbeta$, and $\mathbb{E}_{\bbeta^\star}$ denotes expectation under the sampling distribution determined by $\bbeta^\star$. Once an estimator $\tilde\bbeta$ satisfies this inequality, under the minimax paradigm the statistician may call it minimax optimal and cease attempts to construct a better estimator — because for this class of signals, $\tilde\bbeta$ is already the best estimator in the worst case, up to a multiplicative constant. We refer interested readers to the textbook \cite{polyanskiywu} for a systematic understanding of this paradigm.

The minimax paradigm, although one of the most fundamental research paradigms in mathematical statistics, has received much criticism over the past decades, which we shall not enumerate here; we refer interested readers to the paper of G. Lecu\'e and the author, \cite{lecuesharp2026}. Arising from applications of mathematical statistics in various disciplines, this criticism stems mainly from the pessimism of the minimax paradigm, namely that it focuses on worst-case behavior, whereas in practice $\bbeta^\star$ typically does not lie among these worst-case positions, so that a minimax optimal estimator often does not necessarily exhibit the minimax optimal convergence rate for a given instance of $\bbeta^\star$ in practice. This phenomenon has been reported in many communities, such as quantum physics \cite{ferrie_minimax_2016}, differential privacy \cite{duchi_right_2024}, economics \cite{montiel_olea_decision_nodate}, reinforcement learning \cite{yin_towards_2021}, and statistical physics \cite{abbaras_rademacher_2020}. The discrete probability distribution estimation problem is no exception — whether the minimax paradigm can be used in practice to guide the choice of estimator has long been a controversial question. For example, \cite{orlitsky_competitive_2015,hao_profile_2020} pointed out the following phenomenon: the Laplace estimator, although (nearly) minimax optimal, often performs poorly in practice. This is a central motivation for this work: \textit{can we obtain, for any $\bbeta^\star\in\Delta_p$, the sharp convergence rate of $\mathbb{E}[\KL(\bbeta^\star\|\widehat\bbeta)]$?} The main results of this work answer this question, and hence, for any specific $\bbeta^\star$, provide the estimation error of the Add-$\lambda$ estimator (up to a multiplicative constant), which can in turn explain why its estimation error may be large in that instance.

Below we give some examples to illustrate that the gap between the minimax rate and the instance-wise sharp convergence rate can be very large.

\paragraph{Laplace estimator for Zipf distribution.}
Fix $\alpha>1$ and let
\begin{align*}
\beta_j^\star=\frac{j^{-\alpha}}{\sum_{k=1}^p k^{-\alpha}},\qquad 1\leq j\leq p.
\end{align*}This is known as the Zipf distribution with parameter $\alpha$. The Zipf distribution is a class of probability distributions frequently observed in practice; it appears in, but is not limited to, city populations, the sizes of earthquakes, ..., see \cite[Section 10.2]{polyanskiywu}. Below, taking the Zipf distribution as an example, we examine the expected KL estimation error of the Add-$\lambda$ estimator. The proof of the following corollary may be found in Section~\ref{sec:proof_coro_laplace}.
\begin{Corollary}\label{coro:Laplace}
    There exists $N_0=N_0(\alpha)$ such that for any $N>N_0$ and any $p\geq 2$,
    \begin{align}\label{eq:laplace-decay-examples}
    r^2(V_{J_\star(1)},V_{J_\star(1)^c})\asymp\bE\KL(\bbeta^\star\|\widehat\bbeta^1)\asymp_\alpha\log\frac{N+p}{N+1}.
    \end{align}
\end{Corollary}
Namely, the Zipf distribution attains, up to constants depending on $\alpha$, the worst-case convergence rate of the Laplace estimator ($\lambda=1$), shown by \cite{mourtada2026}:
\begin{align}\label{eq:laplace-worst-case}
\sup_{\bbeta^\star\in\Delta_p}\bE\KL(\bbeta^\star\|\widehat\bbeta^1)=\log\frac{N+p}{N+1}.
\end{align}
Below, we explain for this example, under the condition $p=\lfloor N^\iota\rfloor$, the optimal tuning of the Add-$\lambda$ estimator, as well as the sharp convergence rate of its corresponding expected KL estimation error. We place the computations of \eqref{eq:laplace-decay-examples}, \eqref{eq:oracle-rate-optimality}, and \eqref{eq:zipf-oracle} in Section~\ref{app:examples}.
\paragraph{Oracle tuning for \texorpdfstring{$p=\lfloor N^\iota\rfloor$}{polynomially growing dimension}.}The proof of the following corollary may be found in Section~\ref{sec:proof_coro_Zipf}
\begin{Corollary}\label{coro:Zipf}
    Fix $\iota>0$ and $p=\lfloor N^\iota\rfloor$. There exists $N_0=N_0(\alpha,\iota)$ such that the following statements hold for every $N>N_0$. Let $\lambda^\star$ denote an oracle tuning parameter satisfying
    \begin{align}\label{eq:oracle-rate-optimality}
    r^2(V_{J_\star(\lambda^\star)},V_{J_\star(\lambda^\star)^c})\asymp_{\alpha,\iota}\inf_{\lambda>0}r^2(V_{J_\star(\lambda)},V_{J_\star(\lambda)^c}).
    \end{align}
    For the Zipf distribution with $\alpha>1$, the oracle tuning parameter and its sharp convergence rate satisfy
    \begin{align}\label{eq:zipf-oracle}
    \lambda^\star\asymp_{\alpha,\iota}\begin{cases}1,&0<\iota\leq\frac1\alpha,\\[2pt]N^{1/\alpha-\iota},&\iota>\frac1\alpha,\end{cases}\quad\mbox{and}\quad r^2(V_{J_\star(\lambda^\star)},V_{J_\star(\lambda^\star)^c})\asymp_{\alpha,\iota}\begin{cases}N^{\iota-1},&0<\iota\leq\frac1\alpha,\\[2pt] N^{1/\alpha-1}\log N,&\iota>\frac1\alpha.\end{cases}
    \end{align}
\end{Corollary}
Namely, for a Zipf distribution with parameter $\alpha$, when $p=\lfloor N^\iota\rfloor$, its instance-optimal tuning parameter and instance-optimal sharp convergence rate both exhibit a phase transition at $\iota=1/\alpha$. Since $\alpha$ is unknown to the statistician, we call $\lambda^\star$ an oracle tuning parameter. Constructing a data-dependent parameter selection method that attains the instance-optimal sharp convergence rate for every $\alpha$ is an interesting direction for future research. 

At the end of this subsection, we briefly assess the insight that Corollary~\ref{coro:Zipf} provides for the competitive framework \cite{orlitsky_competitive_2015}. We discuss this using \cite{han_besting_2025} as an example. In \cite{han_besting_2025}, the authors define, as the oracle, the estimator with the smallest expected estimation error among all permutation-invariant estimators (that may depend on $\bbeta^\star$), and show that the expected KL estimation error of the estimator under consideration is close to that of the oracle by establishing an oracle inequality for this estimator relative to the oracle and showing that the residual term of this oracle inequality is small. In particular, the oracle-tuned Add-$\lambda$ estimator considered in Corollary~\ref{coro:Zipf} is permutation-invariant and can serve as the oracle in \cite{han_besting_2025} — hence Corollary~\ref{coro:Zipf} provides, under the Zipf distribution, an upper bound for the oracle expected KL estimation error of the competitive framework, or in the terminology of statistical learning theory, the approximation error of the permutation-invariant class. Although \cite{han_besting_2025} adopts Poisson sampling, whereas this paper adopts multinomial sampling (i.e., $N_j$ is multinomially distributed), so that the conclusions of this paper and those of \cite{han_besting_2025} cannot be strictly compared, Corollary~\ref{coro:Zipf} shows that when $\alpha>3$, the convergence rate (up to a logarithmic factor) of the expected KL estimation error of the oracle-tuned Add-$\lambda$ estimator under multinomial sampling can be significantly faster than $N^{-2/3}$, which is the convergence rate (again, up to logarithmic factor) of the residual term of the oracle inequality established in \cite[Theorem 1.1]{han_besting_2025}. This means that, if we ignore the distinction between multinomial sampling and Poisson sampling, then when $\alpha>3$, the expected KL estimation error of the permutation-invariant oracle can be far smaller than the convergence rate of the residual, and hence in this case the oracle inequality of \cite{han_besting_2025} is insufficient to show that the statistical properties of the estimator under consideration are comparable to those of the oracle. Therefore, Corollary~\ref{coro:Zipf} suggests that the residual term of the oracle inequality in the competitive framework needs to be improved in some special cases. And this is precisely what the FSD series of papers wishes to emphasize: for probability distributions that are of vital importance in the real world, such as the Zipf distribution, a beyond worst-case analysis is needed.

\subsection{Other comments}
In this section, we list the remaining comments.

\subsubsection{On the monotonicity of \texorpdfstring{$r^2(V_{J_\star},V_{J_\star^c})$}{rate} with respect to \texorpdfstring{$\lambda$}{λ}}
With $\lambda$ as the tuning parameter, a natural question is whether one can select the tuning parameter in a data-dependent manner so as to achieve optimal tuning. In \cite{lecueSharpConvergenceRates2025,lecuesharp2026}, by redefining the bias-variance decomposition through the FSD framework, we proved that under fairly general conditions, $\mathrm{BIAS}^{\mathrm{FSD}}$ and $\mathrm{VAR}^{\mathrm{FSD}}$ exhibit opposite monotonicity as $\lambda$ increases, thereby laying the foundation for the use of Lepski's method, \cite{lepskiiProblemAdaptiveEstimation1991}. A natural question is: \textit{for the Add-$\lambda$ estimator, do $\mathrm{BIAS}^{\mathrm{FSD}}$ and $\mathrm{VAR}^{\mathrm{FSD}}$ exhibit opposite monotonicity as $\lambda$ increases, and if so, what are the sufficient (or even necessary) conditions?} The following proposition answers this question affirmatively; its proof may be found in Section~\ref{sec:proof_monotonicity}.
\begin{Proposition}
    \label{prop:monotonicity}
    Fix $N,p\in\bN_+$ and $\bbeta^\star\in\Delta_p$. Then $\lambda\mapsto\mathrm{BIAS}^{\mathrm{FSD}}(\lambda)$ is nondecreasing and $\lambda\mapsto\mathrm{VAR}^{\mathrm{FSD}}(\lambda)$ is nonincreasing on $(0,\infty)$.
\end{Proposition}
How to design an instance-optimal data-dependent parameter selection rule lies beyond the scope of this paper, and we leave it for future work.

\subsubsection{There is no phase transition of \texorpdfstring{$r^2(V_{J_\star(\lambda)},V_{J_\star^c(\lambda)})$}{rate} at \texorpdfstring{$\lambda=1$}{λ=1}}

By examining the definition of $\VAR^{\mathrm{FSD}}$, we find that at $\lambda=1$, $\VAR^{\mathrm{FSD}}$ exhibits a phase transition. However, we emphasize here that although $\VAR^{\mathrm{FSD}}$ exhibits a phase transition, $r^2(V_{J_\star(\lambda)},V_{J_\star^c(\lambda)})$ does not exhibit a phase transition at $\lambda=1$, and hence the Add-$\lambda$ estimator has no phase transition at $\lambda=1$. This can be seen from the following property. The proof of Proposition~\ref{prop:no_phase_transition} may be found in Section~\ref{sec:proof_prop_no_phase_transition}.

\begin{Proposition}\label{prop:no_phase_transition}
For every $N,p\in\mathbb N_+$, $\bbeta^\star\in\Delta_p$, and $\frac12\leq\lambda\leq2$, we have
\begin{align*}
\frac1{32}r^2(V_{J_\star(1)},V_{J_\star(1)^c})\leq r^2(V_{J_\star(\lambda)},V_{J_\star(\lambda)^c})\leq32r^2(V_{J_\star(1)},V_{J_\star(1)^c}),
\end{align*}
where $J_\star(\lambda)=\{1\leq j\leq p:\beta_j^\star>\frac{\lambda}{4(N+\lambda p)}\}$.
\end{Proposition}Together with Theorem~\ref{thm:main}, Proposition~\ref{prop:no_phase_transition} gives a simple but important implication:
\begin{framed}
     for any $N,p\in\mathbb{N}_+$, the (expected) estimation errors of the Krichevsky–Trofimov estimator ($\lambda=\frac{1}{2}$) and the Laplace estimator ($\lambda=1$) are equivalent up to constants for any $\bbeta^\star\in\Delta_p$.
\end{framed}
\begin{Remark}
    The simulations in \cite[Figure 2(d)]{orlitsky_competitive_2015} compare the expected KL estimation error of the Laplace estimator and the Krichevsky–Trofimov estimator for a Zipf distribution with parameter $1.5$, with $p=10000$ and $N$ ranging from $10000$ to $50000$. Although at first glance, the simulation curves therein appear to be at odds with the equivalence up to a multiplicative constant given by Proposition~\ref{prop:no_phase_transition}, their numerical results show that the ratio of the expected KL estimation error of the Laplace estimator to that of the Krichevsky–Trofimov estimator lies approximately between $1.7$ and $2$, which falls within the range of constants allowed by combining Theorem~\ref{thm:main} and Proposition~\ref{prop:no_phase_transition} (which is $[5376^{-1},5376]$). Similar simulation may be found in \cite{han_besting_2025} as well.
\end{Remark}

\subsubsection{The failure of the maximum likelihood estimator}
Although Theorem~\ref{thm:main} holds for $\lambda>0$, we can still let $\lambda\to0^+$ in order to examine the KL estimation error of the maximum likelihood estimator. Fix arbitrary $N,p\in\mathbb{N}_+$ and fix an arbitrary $\bbeta^\star\in\Delta_p$ such that $|\supp(\bbeta^\star)|\geq2$; we have
\begin{align*}
    \lim_{\lambda\to 0^+}\frac{r^2(V_{J_\star},V_{J_\star^c})}{\bigl(\sum_{j=1}^p\beta_j^\star(1-\beta_j^\star)^N\bigr)\log(1/\lambda)}=1,
\end{align*}and in particular, $r^2(V_{J_\star},V_{J_\star^c}) = \Theta(\bE[\sum_{j=1}^p\beta_j^\star\one_{\{N_j=0\}}]\log(\frac{1}{\lambda}))$. Namely: as $\lambda\to 0^+$, the expected KL estimation error diverges to infinity, and the rate of divergence is determined by the expected missing mass. This explains why the MLE fails: even if $N\beta_j^\star$ is large, so that $(1-\beta_j^\star)^N \leq \exp(-N\beta_j^\star)$ is very small, as long as $\beta_j^\star > 0$, the event that this symbol is unobserved can cause the KL estimation error to diverge.

\section{Summary}
Using the Feature Space Decomposition framework, we establish non-asymptotic matching upper and lower bounds for the expected KL estimation error of the Add-$\lambda$ estimator for arbitrary $N,p\in\bN_+$, $\lambda>0$, and $\bbeta^\star\in\Delta_p$, thereby identifying its sharp convergence rate.

Based on this result, we provide a geometric perspective on when the Add-$\lambda$ estimator has a small estimation error: on the estimation subspace $V_{J_\star}$, the angle between $2\sqrt{\hat\bbeta_{J_\star}}$ and $2\sqrt{\bbeta_{J_\star}^\star}$ is small; on the free subspace $V_{J_\star^c}$, the Euclidean norm of $2\sqrt{\hat\bbeta_{J_\star^c}}$ is small.

Taking the Zipf distribution as an example, we prove that the expected KL estimation error of the Laplace estimator is of the same order as its worst-case risk over the entire simplex $\Delta_p$. Nevertheless, when $p=\lfloor N^\iota\rfloor$ with $\iota\geq 1$, the oracle-tuned Add-$\lambda$ estimator still achieves the polynomial convergence rate $N^{1/\alpha-1}\log N$, where $\alpha>1$ is the Zipf parameter. This example may serve as a benchmark for evaluating instance-wise estimation performance and competitive oracle inequalities. Furthermore, we prove that, for arbitrary $N,p\in\bN_+$ and $\bbeta^\star\in\Delta_p$, the expected KL estimation errors of the Krichevsky--Trofimov estimator ($\lambda=1/2$) and the Laplace estimator ($\lambda=1$) are equivalent up to universal multiplicative constants.

Two interesting directions for future research are as follows. The first is to use the FSD framework to establish high-probability upper and lower bounds for the KL estimation error of the Add-$\lambda$ estimator, particularly in the regime $0<\lambda<1$. The second is to develop competitive oracle inequalities with instance-wise sharp remainders through the FSD framework, thereby obtaining risk guarantees beyond worst-case competitive optimality.

\section*{Acknowledgment}
\textbf{LLM Usage.} The author developed the proofs presented in this paper with the assistance of AI-based tools. The AI model used was ChatGPT 6 Astra. The author takes full responsibility for the correctness of the proofs.

\appendix
\section{Proof of Theorem~\texorpdfstring{\ref{thm:main}}{1}}\label{sec:proof_main}

In this section, we prove Theorem~\ref{thm:main}, namely the sharp convergence rate of the expected KL estimation error.

Recall that, for any $\bbeta\in\mathbb{R}^p$, we write $\bbeta_{J_\star} = \Proj_{V_{J_\star}}\bbeta$ and $\bbeta_{J_\star^c}=\Proj_{V_{J_\star^c}}\bbeta$. Hence $\mathbb{R}^p=V_{J_\star}\oplus V_{J_\star^c}$, which we call the Feature Space Decomposition. This gives the following decomposition of the KL estimation error:
\begin{align}\label{eq:decomposition_KL}
    \begin{aligned}
        \KL(\bbeta^\star\|\widehat\bbeta) &= \sum_{j=1}^p\Bigl( \beta_j^\star\log\frac{\beta_j^\star}{\widehat\beta_j} - \beta_j^\star + \widehat \beta_j \Bigr) = \sum_{j\in J_\star}\Bigl( \beta_j^\star\log\frac{\beta_j^\star}{\widehat\beta_j} - \beta_j^\star + \widehat \beta_j \Bigr) + \sum_{j\in J_\star^c}\Bigl( \beta_j^\star\log\frac{\beta_j^\star}{\widehat\beta_j} - \beta_j^\star + \widehat \beta_j \Bigr) \\
        &= \KL(\bbeta_{J_\star}^\star\|\widehat\bbeta_{J_\star}) + \KL(\bbeta_{J_\star^c}^\star\|\widehat\bbeta_{J_\star^c}),
    \end{aligned}
\end{align}where we recall that $\KL(\cdot\|\cdot)$ is the generalized KL divergence defined in \eqref{eq:def_KL}. Below we obtain upper and lower bounds for the KL estimation error on the estimation subspace $V_{J_\star}$ and the free subspace $V_{J_\star^c}$, respectively. To this end, we need the following technical lemma. The proof of the following lemma uses Lemma~\ref{lem:moment_N_j}, Lemma~\ref{lem:integral_log}, Lemma~\ref{lem:identity}, Lemma~\ref{lem:upper_expectation}, and Lemma~\ref{lem:covariance}; we postpone their statements to Section~\ref{sec:auxiliary}.

\begin{Lemma}
    \label{lem:jensen-upper}
    For any $N,p\in\bN_+$, any $j\in\{1,\ldots,p\}$, any $\lambda>0$, and any $\bbeta^\star\in\Delta_p$, let
    \begin{align*}
        \Var_j(\lambda) = \frac{N(\beta_j^\star)^2(1-\beta_j^\star)}{(N\beta_j^\star+\lambda)(N\beta_j^\star+\lambda+1)}+\mathbf1_{\{\lambda<1\}}\beta_j^\star(1-\beta_j^\star)^N\log\bigg(1+\frac{N\beta_j^\star}{\lambda}\bigg).
    \end{align*}Then
    \begin{align}
        \frac1{14}\Var_j(\lambda)\leq\beta_j^\star\big[\log(N\beta_j^\star+\lambda)-\bE\log(N_j+\lambda)\big]\leq12\Var_j(\lambda).\label{eq:jensen-upper}
    \end{align}
\end{Lemma}
\begin{proof}

For $\beta_j^\star\in\{0,1\}$, all terms in \eqref{eq:jensen-upper} vanish. Suppose $0<\beta_j^\star<1$.
\begin{enumerate}
    \item We first prove the upper bound.

    \begin{enumerate}
        \item\label{item:jensen-large-lambda} If $\lambda\geq1$, convexity of $u\mapsto u\log u$ gives $(N\beta_j^\star+\lambda)\log(N\beta_j^\star+\lambda)\leq\bE[(N_j+\lambda)\log(N_j+\lambda)]$, hence
        \begin{align*}
            &(N\beta_j^\star+\lambda)\big[\log(N\beta_j^\star+\lambda)-\bE\log(N_j+\lambda)\big]\\
            &\leq \bE[(N_j+\lambda)\log(N_j+\lambda)] - (N\beta_j^\star +\lambda)\bE[\log(N_j+\lambda)] = \bE[N_j\log(N_j+\lambda)] - N\beta_j^\star\bE[\log(N_j+\lambda)] \\
            & = \bE[N_j\log(N_j+\lambda)] - \bE[N_j]\bE[\log(N_j+\lambda)] =\operatorname{Cov}(N_j,\log(N_j+\lambda))\leq\frac{2N\beta_j^\star(1-\beta_j^\star)}{N\beta_j^\star+\lambda},
        \end{align*}where the last inequality is due to Lemma~\ref{lem:covariance}.
        As $N\beta_j^\star+\lambda\geq1$, this proves the upper bound in \eqref{eq:jensen-upper}, with coefficient $4$.
        \item\label{item:jensen-small-lambda} If $0<\lambda<1$, since $\bE\bigl[\frac{N_j-N\beta_j^\star}{N\beta_j^\star+\lambda}\bigr]=(N\beta_j^\star+\lambda)^{-1}(\bE[N_j]-N\beta_j^\star)=0$, we obtain that
        \begin{align}\label{eq:log_exp}
            \begin{aligned}
                &\log(N\beta_j^\star+\lambda)-\bE\log(N_j+\lambda) = \bE\Bigl(\log(N\beta_j^\star+\lambda)-\log(N_j+\lambda)\Bigr) = \bE\log\Bigl(\frac{N\beta_j^\star+\lambda}{N_j+\lambda}\Bigr)\\
                &=\bE\log\Bigl(\frac{N\beta_j^\star+\lambda}{N_j+\lambda}\Bigr) + \bE\Bigl[\frac{N_j-N\beta_j^\star}{N\beta_j^\star+\lambda}\Bigr] = \bE\Bigl[ \frac{N_j-N\beta_j^\star}{N\beta_j^\star+\lambda} - \log\Bigl(\frac{N_j+\lambda}{N\beta_j^\star+\lambda}\Bigr) \Bigr].
            \end{aligned}
        \end{align}By $\log(u)\leq u-1$ for all $u>0$, the random variable $\frac{N_j-N\beta_j^\star}{N\beta_j^\star+\lambda} - \log\bigl(\frac{N_j+\lambda}{N\beta_j^\star+\lambda}\bigr)$ is non-negative. We consider two cases.
        \begin{enumerate}
            \item \label{item:N_j_0} When $N_j=0$. Then
            \begin{align*}
                &\frac{N_j-N\beta_j^\star}{N\beta_j^\star+\lambda} - \log\Bigl(\frac{N_j+\lambda}{N\beta_j^\star+\lambda}\Bigr) = \log\Bigl(1+\frac{N\beta_j^\star}{\lambda}\Bigr) - \frac{N\beta_j^\star}{N\beta_j^\star+\lambda} \leq \log\Bigl(1+\frac{N\beta_j^\star}{\lambda}\Bigr).
            \end{align*}Combining with $\bP(N_j=0)=(1-\beta_j^\star)^N$, we obtain that
            \begin{align*}
                \bP(N_j=0)\biggl(\frac{N_j-N\beta_j^\star}{N\beta_j^\star+\lambda} - \log\Bigl(\frac{N_j+\lambda}{N\beta_j^\star+\lambda}\Bigr)\biggr)\bigg|_{N_j=0} \leq (1-\beta_j^\star)^N\log\Bigl(1+\frac{N\beta_j^\star}{\lambda}\Bigr).
            \end{align*}
            \item When $N_j\geq 1$. By the elementary inequality $u-1-\log(u)\leq \frac{(u-1)^2}{u}$ for all $u>0$ applied to $u=\frac{N_j+\lambda}{N\beta_j^\star+\lambda}$, we obtain
            \begin{align*}
                &\frac{N_j-N\beta_j^\star}{N\beta_j^\star+\lambda} - \log\Bigl(\frac{N_j+\lambda}{N\beta_j^\star+\lambda}\Bigr) \leq \frac{\Bigl(\frac{N_j+\lambda}{N\beta_j^\star+\lambda}-1\Bigr)^2}{\frac{N_j+\lambda}{N\beta_j^\star+\lambda}} = \frac{(N_j-N\beta_j^\star)^2}{(N\beta_j^\star+\lambda)(N_j+\lambda)} \leq \frac{2(N_j-N\beta_j^\star)^2}{(N\beta_j^\star+\lambda)(N_j+1)},
            \end{align*}where we have used the fact that on the random event $\{N_j\geq 1\}$, $N_j+\lambda\geq N_j\geq \frac{1}{2}(N_j+1)$ in the last inequality. Taking expectations on both sides together with Lemma~\ref{lem:upper_expectation} yields
            \begin{align*}
                &\bE\biggl[\biggl(\frac{N_j-N\beta_j^\star}{N\beta_j^\star+\lambda} - \log\Bigl(\frac{N_j+\lambda}{N\beta_j^\star+\lambda}\Bigr)\biggr)\one_{\{N_j\geq 1\}}\biggr] \leq \bE\biggl[ \frac{2(N_j-N\beta_j^\star)^2}{(N\beta_j^\star+\lambda)(N_j+1)}\one_{\{N_j\geq 1\} } \biggr]\\
                &\leq \bE\biggl[ \frac{2(N_j-N\beta_j^\star)^2}{(N\beta_j^\star+\lambda)(N_j+1)}\biggr]\leq \frac{6N\beta_j^\star(1-\beta_j^\star)}{(N\beta_j^\star+\lambda)(N\beta_j^\star+1)} \leq \frac{12N\beta_j^\star(1-\beta_j^\star)}{(N\beta_j^\star+\lambda)(N\beta_j^\star+\lambda+1)},
            \end{align*}where we have used the fact that $\lambda<1$ and hence $N\beta_j^\star+\lambda+1\leq 2(N\beta_j^\star+1)$ in the last inequality.
        \end{enumerate}Combining the two cases and multiplying both sides by $\beta_j^\star$ gives the upper bound of \eqref{eq:jensen-upper} when $\lambda<1$.
        
    Combining item~\emph{\ref{item:jensen-large-lambda}} and item~\emph{\ref{item:jensen-small-lambda}} gives the upper bound of \eqref{eq:jensen-upper} for all $\lambda$.
    \end{enumerate}
    \item Below we prove the lower bound. By Lemma~\ref{lem:integral_log},
    \begin{align}\tag{\ref{eq:resolvent}}
    \log(N\beta_j^\star+\lambda)-\bE\log(N_j+\lambda)=\int_\lambda^\infty\frac{\bE\big[\frac{(N_j-N\beta_j^\star)^2}{N_j+s}\big]}{(N\beta_j^\star+s)^2}\,\mathrm ds.
    \end{align} Below we examine the lower bound for $\mathbb{E}\big[\frac{(N_j-N\beta_j^\star)^2}{N_j+s}\big]$. For any $s>0$, by Cauchy's inequality,
    \begin{align*}
        \Bigl(\bE\bigl[( N_j - N\beta_j^\star)^2\bigr]\Bigr)^2 &= \Bigl( \bE\Bigl[ \frac{N_j-N\beta_j^\star}{\sqrt{N_j+s}}\cdot (N_j-N\beta_j^\star)\sqrt{N_j+s} \Bigr] \Bigr)^2\\
        &\leq \bE\Bigl[\frac{(N_j-N\beta_j^\star)^2}{N_j+s}\Bigr] \bE\Bigl[(N_j-N\beta_j^\star)^2(N_j+s)\Bigr].
    \end{align*}Now by Lemma~\ref{lem:moment_N_j}, the above displayed inequality is equivalent to
    \begin{align*}
        \Bigl(N\beta_j^\star(1-\beta_j^\star)\Bigr)^2 \leq \bE\Bigl[\frac{(N_j-N\beta_j^\star)^2}{N_j+s}\Bigr] \Bigl( N\beta_j^\star(1-\beta_j^\star)(N\beta_j^\star+s+1-2\beta_j^\star)\Bigr).
    \end{align*}Since $0<\beta_j^\star<1$ and $s>0$, the above displayed inequality is further equivalent to
    \begin{align*}
        \bE\Bigl[\frac{(N_j-N\beta_j^\star)^2}{N_j+s}\Bigr] &\geq \frac{ \Bigl(N\beta_j^\star(1-\beta_j^\star)\Bigr)^2 }{ N\beta_j^\star(1-\beta_j^\star)(N\beta_j^\star+s+1-2\beta_j^\star) } \geq \frac{N\beta_j^\star(1-\beta_j^\star)}{N\beta_j^\star+s+1}.
    \end{align*}Substituting this into \eqref{eq:resolvent},
    \begin{align}
        \log(N\beta_j^\star+\lambda)-\bE\log(N_j+\lambda)\geq\int_\lambda^\infty\frac{N\beta_j^\star(1-\beta_j^\star)}{(N\beta_j^\star+s)^2(N\beta_j^\star+s+1)}\,\mathrm ds\geq\frac{N\beta_j^\star(1-\beta_j^\star)}{2(N\beta_j^\star+\lambda)(N\beta_j^\star+\lambda+1)},\label{eq:quadratic-lower}
    \end{align}where the last inequality follows from $-\frac{\mathrm d}{\mathrm dx}\frac1{2x(x+1)} = \frac{2x+1}{2x^2(x+1)^2} = \frac{1}{x^2(x+1)}\frac{2x+1}{2(x+1)} \leq\frac1{x^2(x+1)}$ for $x>0$. This yields the lower bound of \eqref{eq:jensen-upper} when $\lambda\geq 1$. To finish the proof of the lower bound of \eqref{eq:jensen-upper}, we are left to prove that when $0<\lambda<1$,
    \begin{align}\label{eq:objective_lambda_small}
        (1-\beta_j^\star)^N\log\Bigl(1+\frac{N\beta_j^\star}{\lambda}\Bigr) \leq 12 \Bigl( \log(N\beta_j^\star+\lambda) - \bE\bigl[\log(N_j+\lambda)\bigr] \Bigr).
    \end{align}This is in fact already done in item~\emph{\ref{item:N_j_0}}. Indeed, by \eqref{eq:log_exp},
    \begin{align*}
        &\log(N\beta_j^\star+\lambda)-\bE\log(N_j+\lambda) \geq \bE\Bigl[ \Bigl(\frac{N_j-N\beta_j^\star}{N\beta_j^\star+\lambda} - \log\Bigl(\frac{N_j+\lambda}{N\beta_j^\star+\lambda}\Bigr) \Bigr)\one_{\{N_j=0\}}\Bigr]\\
        &=\bP(N_j=0)\biggl(\frac{N_j-N\beta_j^\star}{N\beta_j^\star+\lambda} - \log\Bigl(\frac{N_j+\lambda}{N\beta_j^\star+\lambda}\Bigr)\biggr)\bigg|_{N_j=0} = (1-\beta_j^\star)^N\Bigl(\log\bigl(1+\frac{N\beta_j^\star}{\lambda}\bigr) - \frac{N\beta_j^\star}{N\beta_j^\star+\lambda}\Bigr).
    \end{align*}We consider two cases.
    \begin{enumerate}
        \item When $N\beta_j^\star\geq\lambda$. From the elementary inequality $\log(1+u)-\frac{u}{1+u}\geq\frac{1}{4}\log(1+u)$ for all $u\geq 1$, applied to $u=\frac{N\beta_j^\star}{\lambda}$, we obtain
        \begin{align*}
            &\log(N\beta_j^\star+\lambda)-\bE\log(N_j+\lambda) \geq \frac{1}{4}(1-\beta_j^\star)^N\log\Bigl(1+\frac{N\beta_j^\star}{\lambda}\Bigr).
        \end{align*}
        \item When $N\beta_j^\star<\lambda$. By $\lambda<1$, elementary inequality $\log(1+u)\leq u$, and $(1-\beta_j^\star)^N\leq 1-\beta_j^\star$,
        \begin{align*}
            (1-\beta_j^\star)^N \log\bigl(1+\frac{N\beta_j^\star}{\lambda}\bigr)&\leq \frac{N\beta_j^\star(1-\beta_j^\star)}{\lambda} \leq \frac{6N\beta_j^\star(1-\beta_j^\star)}{(N\beta_j^\star+\lambda)(N\beta_j^\star+\lambda+1)}\\
            &\leq 12\Bigl(\log(N\beta_j^\star+\lambda)-\bE[\log(N_j+\lambda)]\Bigr),
        \end{align*}where we have used $N\beta_j^\star+\lambda<2\lambda$, $N\beta_j^\star+\lambda+1<3$, and \eqref{eq:quadratic-lower}.
    \end{enumerate}Combining the two cases, for any $0<\lambda<1$, \eqref{eq:objective_lambda_small} holds. Multiplying both sides by $\beta_j^\star$ of \eqref{eq:quadratic-lower}, \eqref{eq:objective_lambda_small}, and summing, the lower bound of \eqref{eq:jensen-upper} is verified.
\end{enumerate}
\end{proof}

\subsection{Estimation error on \texorpdfstring{$V_{J_\star}$}{Estimation subspace}}

\begin{Lemma}
    \label{lem:estimation-upper}
    For any $\lambda>0$, any $N,p\in\bN_+$, and any $\bbeta^\star\in\Delta_p$,
    \begin{align*}
        \begin{aligned}
            &\bE\KL(\bbeta_{J_\star}^\star\|\widehat{\bbeta}_{J_\star})=\mathrm{Bias}_{J_\star}+\sum_{j\in J_\star}\beta_j^\star\big[\log(N\beta_j^\star+\lambda)-\bE\log(N_j+\lambda)\big]\asymp\mathrm{Bias}_{J_\star}+\mathrm{VAR}^{\mathrm{FSD}},
        \end{aligned}
    \end{align*}where the absolute constants may be taken to be $\frac{1}{14}$ and $12$.
\end{Lemma}
\begin{proof}

    We first notice that by definition, $\widehat\beta_j = \frac{N_j+\lambda}{N+\lambda p}$, hence
    \begin{align*}
        &\KL(\bbeta_{J_\star}^\star\|\widehat\bbeta_{J_\star}) = \sum_{j\in J_\star}\Bigl( \beta_j^\star\log\frac{\beta_j^\star}{\widehat\beta_j} - \beta_j^\star + \widehat\beta_j \Bigr) = \sum_{j\in J_\star}\Bigl( \beta_j^\star\log\frac{(N+\lambda p)\beta_j^\star}{N_j+\lambda} - \beta_j^\star + \frac{N_j+\lambda}{N+\lambda p} \Bigr).
    \end{align*}Now using $\beta_j^\star\log\frac{(N+\lambda p)\beta_j^\star}{N_j+\lambda} = \beta_j^\star\log( \frac{(N+\lambda p)\beta_j^\star}{N\beta_j^\star+\lambda}\cdot \frac{N\beta_j^\star+\lambda}{N_j+\lambda}) = \beta_j^\star\log\frac{(N+\lambda p)\beta_j^\star}{N\beta_j^\star+\lambda} + \beta_j^\star\log\frac{N\beta_j^\star+\lambda}{N_j+\lambda}$; together with $\frac{N_j+\lambda}{N+\lambda p} = \frac{N\beta_j^\star+\lambda}{N+\lambda p} + \frac{N_j - N\beta_j^\star}{N+\lambda p}$, we obtain that
    \begin{align*}
        \KL(\bbeta_{J_\star}^\star\|\widehat\bbeta_{J_\star}) &= \biggl(\sum_{j\in J_\star} \Bigl[\beta_j^\star\log\frac{(N+\lambda p)\beta_j^\star}{N\beta_j^\star+\lambda} + \frac{N\beta_j^\star +\lambda}{N+\lambda p} - \beta_j^\star\Bigr]\biggr) + \biggl( \sum_{j\in J_\star} \Bigl[\beta_j^\star\log\frac{N\beta_j^\star+\lambda}{N_j+\lambda} + \frac{N_j - N\beta_j^\star}{N+\lambda p}\Bigr]\biggr)\\
        &=\mathrm{Bias}_{J_\star} + \biggl( \sum_{j\in J_\star} \Bigl[\beta_j^\star\log\frac{N\beta_j^\star+\lambda}{N_j+\lambda} + \frac{N_j - N\beta_j^\star}{N+\lambda p}\Bigr]\biggr).
    \end{align*}Taking expectations on both sides, together with the observation that $\bE[N_j]=N\beta_j^\star$, we obtain that
    \begin{align*}
        \bE\KL(\bbeta_{J_\star}^\star\|\widehat\bbeta_{J_\star}) &= \Bias_{J_\star} + \sum_{j\in J_\star} \beta_j^\star\bE\log\frac{N\beta_j^\star+\lambda}{N_j+\lambda} = \Bias_{J_\star} + \sum_{j\in J_\star} \beta_j^\star\Bigl(\log(N\beta_j^\star+\lambda)-\bE\log(N_j+\lambda)\Bigr).
    \end{align*}Lemma~\ref{lem:estimation-upper} follows by applying Lemma~\ref{lem:jensen-upper} to control $\log(N\beta_j^\star+\lambda)-\bE\log(N_j+\lambda)$.
\end{proof}

\subsection{Energy of \texorpdfstring{$\widehat\bbeta_{J_\star^c}$}{free part of the estimator}}

Below we examine $\KL(\bbeta_{J_\star^c}^\star\|\widehat\bbeta_{J_\star^c})$. We first collect two important consequences of the definition of $J_\star$ (see \eqref{eq:decomposition}). Recall that from \eqref{eq:decomposition}, there holds
\begin{align}\label{eq:filter}
    \forall j\in J_\star^c,\qquad \beta_j^\star\leq\frac{\lambda}{4(N+\lambda p)} \leq \frac{\lambda}{4},
\end{align}where we have used that $N\geq 1$ and $\lambda>0$. Thus, $\widehat\bbeta$ has a filtering effect: once the statistician specifies $\lambda$, the coordinates assigned to the free subspace satisfy $\beta_j^\star\leq\frac{\lambda}{4(N+\lambda p)}$. In particular,
\begin{align}\label{eq:separation}
    \forall j\in J_\star^c,\qquad \widehat\beta_j = \frac{N_j+\lambda}{N+\lambda p}\geq\frac{\lambda}{N+\lambda p}\geq 4\beta_j^\star.
\end{align}From \eqref{eq:separation}, we can intuitively see why $\widehat\bbeta$ does not estimate these coordinates: for these coordinates, $\widehat\beta_j$ is at least four times $\beta_j^\star$.
\begin{Lemma}\label{lem:free-upper}
    For any $\lambda>0$, any $N,p\in\bN_+$, and any $\bbeta^\star\in\Delta_p$,
    \begin{align*}
        \frac{3-\log(4)}{4}\Bias_{J_\star^c}\leq\bE[\KL(\bbeta_{J_\star^c}^\star\|\widehat\bbeta_{J_\star^c})]\leq\Bias_{J_\star^c}.
    \end{align*}
    Moreover, almost surely,
    \begin{align*}
        \frac{3-\log(4)}{4}\sum_{j\in J_\star^c}\widehat\beta_j \leq \KL(\bbeta_{J_\star^c}^\star\|\widehat\bbeta_{J_\star^c}) \leq \sum_{j\in J_\star^c}\widehat\beta_j.
    \end{align*}
\end{Lemma}
\begin{proof}
    By \eqref{eq:separation}, $0\leq\frac{\beta_j^\star}{\widehat\beta_j}\leq\frac14$ for every $j\in J_\star^c$. Since $u\mapsto1-u+u\log u$ is nonincreasing on $[0,\frac14]$, with $0\log0=0$,
    \begin{align}\label{eq:each_coordinate}
        \frac{3-\log(4)}{4}\widehat\beta_j\leq\widehat\beta_j\bigg(1-\frac{\beta_j^\star}{\widehat\beta_j}+\frac{\beta_j^\star}{\widehat\beta_j}\log\frac{\beta_j^\star}{\widehat\beta_j}\bigg)=\beta_j^\star\log\frac{\beta_j^\star}{\widehat\beta_j}-\beta_j^\star+\widehat\beta_j\leq\widehat\beta_j.
    \end{align}
    Summing over $j\in J_\star^c$ proves the almost sure bounds. Its lower bound and $\widehat\beta_j\geq\frac\lambda{N+\lambda p}$ give, almost surely,
    \begin{align*}
        \KL(\bbeta_{J_\star^c}^\star\|\widehat\bbeta_{J_\star^c})\geq\frac{3-\log(4)}{4}\frac{\lambda|J_\star^c|}{N+\lambda p}=\frac{3-\log(4)}{4}\Bias_{J_\star^c}.
    \end{align*}
    For the expected upper bound, use the definition of $\hat\bbeta$, $N_j\geq 0$, $\widehat\beta_j\geq\frac\lambda{N+\lambda p}$, $\bE\widehat\beta_j=\frac{N\beta_j^\star+\lambda}{N+\lambda p}$, $\Bias_{J_\star^c}=\frac{\lambda|J_\star^c|}{N+\lambda p}$, and \eqref{eq:filter} to obtain
    \begin{align*}
        &\bE[\KL(\bbeta_{J_\star^c}^\star\|\widehat\bbeta_{J_\star^c})] = \bE\sum_{j\in J_\star^c}\biggl( \beta_j^\star\log\frac{(N+\lambda p)\beta_j^\star}{N_j+\lambda} - \beta_j^\star + \frac{N_j+\lambda}{N+\lambda p} \biggr) \leq\sum_{j\in J_\star^c}\biggl(\beta_j^\star\log\frac{(N+\lambda p)\beta_j^\star}{\lambda}-\beta_j^\star+\frac{N\beta_j^\star+\lambda}{N+\lambda p}\biggr)\\
        & =\sum_{j\in J_\star^c}\biggl(\beta_j^\star\log\frac{(N+\lambda p)\beta_j^\star}{\lambda} + \frac{\lambda}{N+\lambda p} - \frac{\lambda p\beta_j^\star}{N+\lambda p}  \biggr) =\Bias_{J_\star^c}+\sum_{j\in J_\star^c}\beta_j^\star\bigg[\log\frac{(N+\lambda p)\beta_j^\star}{\lambda}-\frac{\lambda p}{N+\lambda p}\bigg]\leq\Bias_{J_\star^c}.
    \end{align*}
\end{proof}

\subsection{End of the proof of Theorem~\texorpdfstring{\ref{thm:main}}{expectation sharp convergence rate}}

Taking expectations in \eqref{eq:decomposition_KL} and applying Lemmas~\ref{lem:estimation-upper} and \ref{lem:free-upper}, together with the definition of $r^2(V_{J_\star},V_{J_\star^c})$, completes the proof of Theorem~\ref{thm:main}.

\section{Calculations for the examples}\label{app:examples}

Throughout this appendix, $\alpha>1$ and $\beta_j^\star=j^{-\alpha}/\sum_{k=1}^p k^{-\alpha}$ for $1\leq j\leq p$. Since $1\leq\sum_{k=1}^p k^{-\alpha}\leq1+\int_1^\infty t^{-\alpha}\,\mathrm dt=\frac\alpha{\alpha-1}$, we have:
\begin{align*}
    \forall 1\leq j\leq p,\qquad \frac{\alpha-1}{\alpha}j^{-\alpha}\leq\beta_j^\star\leq j^{-\alpha}.
\end{align*}
Also, $1-\beta_j^\star\geq1-\beta_1^\star\geq\frac{2^{-\alpha}}{1+2^{-\alpha}}\geq2^{-\alpha-1}$ whenever $p\geq2$. Dependence of comparison constants on $\alpha,\iota,c,C$ is indicated by subscripts. In statements involving $p=\lfloor N^\iota\rfloor$, $\iota>0$ is fixed.

\subsection{Proof of Corollary~\texorpdfstring{\ref{coro:Laplace}}{laplace}}\label{sec:proof_coro_laplace}

Fix $C_\alpha=2+\frac{\log(1+\alpha/(\alpha-1))}{\alpha-1}+\frac{\alpha}{(\alpha-1)^2}$ and $N_0=\bigl\lceil\max\{2,\frac{2^\alpha\alpha}{\alpha-1},(\frac{8\alpha C_\alpha}{(\alpha-1)\log2})^{\alpha/(\alpha-1)}\}\bigr\rceil$, and suppose $N>N_0$.
We calculate the three quantities in \eqref{eq:fsd-components} at $\lambda=1$. Recall that
\begin{align}\tag{\ref{eq:fsd-components}}
    \begin{aligned}&\mathrm{Bias}_{J_\star}=\sum_{j\in J_\star}\biggl[\beta_j^\star\log\frac{(N+\lambda p)\beta_j^\star}{N\beta_j^\star+\lambda}+\frac{N\beta_j^\star+\lambda}{N+\lambda p}-\beta_j^\star\biggr],\quad \mathrm{Bias}_{J_\star^c}=\frac{\lambda|J_\star^c|}{N+\lambda p},\quad\mbox{ and }\\ &\\ &\mathrm{VAR}^{\mathrm{FSD}}=\sum_{j\in J_\star}\biggl[\frac{N(\beta_j^\star)^2(1-\beta_j^\star)}{(N\beta_j^\star+\lambda)(N\beta_j^\star+\lambda+1)}+\mathbf1_{\{\lambda<1\}}\beta_j^\star(1-\beta_j^\star)^N\log\biggl(1+\frac{N\beta_j^\star}{\lambda}\biggr)\biggr].\end{aligned}
\end{align}
The upper bound follows from \cite{mourtada2026} and Theorem~\ref{thm:main}, so it remains to bound $r^2(V_{J_\star(1)},V_{J_\star^c(1)})$ from below.
\paragraph{Calculation of \texorpdfstring{$J_\star(1)$}{the estimation index set}.}
We first calculate $J_\star(\lambda)$. Recall that
\begin{align}\tag{\ref{eq:decomposition}}
    J_\star(\lambda)=\bigg\{1\leq j\leq p:\beta_j^\star>\frac\lambda{4(N+\lambda p)}\bigg\}.
\end{align}
Substituting the probabilities into \eqref{eq:decomposition}, the inequality defining this set becomes $j^{-\alpha}>\frac\lambda{4(N+\lambda p)}\sum_{k=1}^p k^{-\alpha}$. Taking reciprocals and then the positive $\alpha$-th root gives
\begin{equation}\label{eq:examples-index-set}
    \begin{aligned}J_\star(\lambda)&=\biggl\{1\leq j\leq p:j<\biggl(\frac{4(N+\lambda p)}{\lambda\sum_{k=1}^p k^{-\alpha}}\biggr)^{1/\alpha}\biggr\},\\ |J_\star(\lambda)|&=\min\biggl\{p,\biggl\lceil\biggl(\frac{4(N+\lambda p)}{\lambda\sum_{k=1}^p k^{-\alpha}}\biggr)^{1/\alpha}\biggr\rceil-1\biggr\},\qquad |J_\star(1)|\asymp_\alpha\min\{p,(N+p)^{1/\alpha}\}.\end{aligned}
\end{equation}
The ceiling and the subtraction by one account for the strict inequality in the definition of $J_\star(\lambda)$. The last comparison follows from $1\leq\sum_{k=1}^p k^{-\alpha}\leq\frac\alpha{\alpha-1}$ and $x/2\leq\lceil x\rceil-1\leq x$ for $x\geq2$.

Below, we compute lower bounds for $\Bias_{J_\star(1)}$, $\Bias_{J_\star^c(1)}$, and $\VAR^\FSD(1)$ in three cases, respectively.
\begin{enumerate}
    \item When $p\geq N$, we show that $\Bias_{J_\star(1)}\gtrsim_\alpha\log\Bigl(1+\frac{p}{N}\Bigr)$.
    \item When $2(8N)^{1/\alpha}\leq p\leq N$, we show that $\Bias_{J_\star^c(1)}\gtrsim_\alpha\log\Bigl(1+\frac{p}{N}\Bigr)$.
    \item When $p\leq 2(8N)^{1/\alpha}$, we show that $\VAR^\FSD(1)\gtrsim_\alpha\log\Bigl(1+\frac{p}{N}\Bigr)$.
\end{enumerate}The above three cases cover all possible relationships between $N$ and $p$.

\paragraph{Calculation of \texorpdfstring{$\mathrm{Bias}_{J_\star(1)}$}{the estimation subspace bias}.}
We establish the lower bound on $\Bias_{J_\star(1)}$ only when $p\geq N$. Recall that
\begin{align}\tag{\ref{eq:fsd-components}}
    \begin{aligned}
        \mathrm{Bias}_{J_\star(1)}&=\sum_{j\in J_\star(1)}\biggl(\beta_j^\star\log\frac{(N+ p)\beta_j^\star}{N\beta_j^\star+1}+\frac{N\beta_j^\star+1}{N+ p}-\beta_j^\star\biggr) =\sum_{j\in J_\star(1)}\biggl[\beta_j^\star\log\Bigl(\frac{1+\frac{p}{N}}{1+\frac{1}{N\beta_j^\star}}\Bigr)+\frac{1-p\beta_j^\star}{N+p}\biggr]\\
        &=\sum_{j\in J_\star(1)}\beta_j^\star\biggl(\log\Bigl(1+\frac{p}{N}\Bigr) - \frac{p}{N+p}\biggr) - \sum_{j\in J_\star(1)}\beta_j^\star\log\Bigl(1+\frac{1}{N\beta_j^\star}\Bigr) + \frac{|J_\star(1)|}{N+p}.
    \end{aligned}
\end{align}
\begin{enumerate}
    \item We first estimate $\sum_{j\in J_\star(1)}\beta_j^\star\log(1+\frac1{N\beta_j^\star})$ from above.

    For $j\leq\lceil N^{1/\alpha}\rceil$, the inequality $\log(1+x)\leq x$ bounds each summand by $1/N$, so their sum is at most $2N^{1/\alpha-1}$. For $j>\lceil N^{1/\alpha}\rceil$, the bounds on $\beta_j^\star$ give $\beta_j^\star\log(1+\frac1{N\beta_j^\star})\leq j^{-\alpha}\log(1+\frac\alpha{\alpha-1}\frac{j^\alpha}{N})$. The function $t\mapsto t^{-\alpha}\log(1+\frac\alpha{\alpha-1}\frac{t^\alpha}{N})$ is decreasing: after differentiation, its derivative has the sign of $\frac{x}{1+x}-\log(1+x)<0$ for $x>0$. Hence
    \begin{align*}
        j^{-\alpha}\log\Bigl(1+\frac{\alpha j^\alpha}{(\alpha-1)N}\Bigr) \leq \int_{j-1}^j t^{-\alpha}\log\Bigl(1+\frac{\alpha t^\alpha}{(\alpha-1)N}\Bigr)\mathrm{d}t.
    \end{align*}Summing over $j>\lceil N^{1/\alpha}\rceil$, we obtain $$\sum_{j>\lceil N^{1/\alpha}\rceil}j^{-\alpha}\log(1+\frac\alpha{\alpha-1}\frac{j^\alpha}{N})\leq N^{1/\alpha-1}\int_1^\infty u^{-\alpha}\log(1+\frac\alpha{\alpha-1}u^\alpha)\,\mathrm du.$$ Since $u\geq1$ implies $1+\frac\alpha{\alpha-1}u^\alpha\leq(1+\frac\alpha{\alpha-1})u^\alpha$, this integral is at most $\frac{\log(1+\alpha/(\alpha-1))}{\alpha-1}+\frac\alpha{(\alpha-1)^2}$. Indeed, $\int_1^\infty u^{-\alpha}\,\mathrm du=\frac1{\alpha-1}$ and integration by parts gives $\int_1^\infty u^{-\alpha}\log u\,\mathrm du=\frac1{(\alpha-1)^2}$. Consequently,
    \begin{align*}
        \sum_{j\in J_\star(1)}\beta_j^\star\log(1+\frac1{N\beta_j^\star})\leq \biggl(2+\frac{\log(1+\alpha/(\alpha-1))}{\alpha-1}+\frac\alpha{(\alpha-1)^2}\biggr)N^{1/\alpha-1},
    \end{align*}As a consequence,
    \begin{align*}
        &\sum_{j\in J_\star(1)}\beta_j^\star\log\Bigl(1+\frac1{N\beta_j^\star}\Bigr) \leq C_\alpha N^{\frac{1}{\alpha}-1}.
    \end{align*}

    \item We then bound $\log\Bigl(1+\frac{p}{N}\Bigr) - \frac{p}{N+p}$ from below.

    For $p\geq N$, $\log(1+\frac pN)-\frac p{N+p}\geq\frac14\log(1+\frac pN)$. To verify this, the difference $\frac34\log(1+x)-\frac{x}{1+x}$ is positive at $x=1$, and its derivative is $\frac{3x-1}{4(1+x)^2}>0$ for $x\geq1$.

    \item We next bound the signal mass on $J_\star(1)$ from below.

    By the choice of $N_0$, $1\in J_\star(1)$, hence $\sum_{j\in J_\star(1)}\beta_j^\star\geq\beta_1^\star\geq\frac{\alpha-1}{\alpha}$. 
\end{enumerate}
Therefore
\begin{align*}
    \mathrm{Bias}_{J_\star(1)}\geq\frac{\alpha-1}{4\alpha}\log(1+\frac pN) - C_\alpha N^{\frac{1}{\alpha}-1} > \frac{\alpha-1}{8\alpha}\log\Bigl(1+\frac{p}{N}\Bigr),\qquad\mbox{ when }\qquad p\geq N,
\end{align*}by the choice of $N_0$.

\paragraph{Calculation of \texorpdfstring{$\mathrm{Bias}_{J_\star(1)^c}$}{the free subspace bias}.} We establish the lower bound on $\Bias_{J_\star^c(1)}$ only when $2(8N)^{1/\alpha}\leq p\leq N.$ Recall that
\begin{align}\tag{\ref{eq:fsd-components}}
    \mathrm{Bias}_{J_\star^c(1)}=\frac{|J_\star^c(1)|}{N+ p}.
\end{align}
Then \eqref{eq:examples-index-set} gives $|J_\star(1)|\leq[4(N+p)]^{1/\alpha}\leq(8N)^{1/\alpha}\leq p/2$, and $N+p\leq2N$ then gives $(p-|J_\star(1)|)/(N+p)\geq p/(4N)$, hence
\begin{align*}
    \mathrm{Bias}_{J_\star(1)^c}\geq\frac p{4N}\geq\frac{1}{4}\log\Bigl(1+\frac{p}{N}\Bigr),\qquad\mbox{ when }\qquad 2(8N)^{1/\alpha}\leq p\leq N.
\end{align*}

\paragraph{Calculation of \texorpdfstring{$\mathrm{VAR}^{\mathrm{FSD}}(1)$}{the FSD variance}.}
At $\lambda=1$ the indicator in \eqref{eq:fsd-components} vanishes, so we calculate
\begin{equation}\label{eq:examples-laplace-variance}
    \mathrm{VAR}^{\mathrm{FSD}}(1)=\frac1N\sum_{j\in J_\star(1)}(1-\beta_j^\star)\frac{(N\beta_j^\star)^2}{(N\beta_j^\star+1)(N\beta_j^\star+2)}.
\end{equation}We only establish its lower bound when $p\leq N$ and $p\leq 2(8N)^{1/\alpha}$. Using
\begin{align}\tag{\ref{eq:examples-index-set}}
    |J_\star(1)|&=\min\biggl\{p,\biggl\lceil\biggl(\frac{4(N+ p)}{\sum_{k=1}^p k^{-\alpha}}\biggr)^{1/\alpha}\biggr\rceil-1\biggr\},
\end{align}
together with $\sum_{k=1}^pk^{-\alpha}\leq\frac{\alpha}{\alpha-1}$, we obtain
\begin{align*}
    |J_\star(1)|\geq \min\Bigl\{p,\Bigl\lfloor\bigl( \frac{(\alpha-1)N}{\alpha} \bigr)^{1/\alpha}\Bigr\rfloor\Bigr\},
\end{align*}For each $1\leq j\leq \min\{p,\lfloor\bigl( \frac{(\alpha-1)N}{\alpha} \bigr)^{1/\alpha}\rfloor\}$, $N\beta_j^\star\geq\frac{\alpha-1}{\alpha}\frac{N}{j^\alpha}\geq 1$. Hence $N\beta_j^\star+1\leq 2N\beta_j^\star$, and $N\beta_j^\star+2\leq 3N\beta_j^\star$, and consequently,
\begin{align*}
    \VAR^\FSD(1) \geq \frac{2^{-\alpha-1}}{6N}\min\Bigl\{p,\Bigl\lfloor\bigl( \frac{(\alpha-1)N}{\alpha} \bigr)^{1/\alpha}\Bigr\rfloor\Bigr\}.
\end{align*}Now since $N\geq \frac{\alpha}{\alpha-1}2^\alpha$ and $p\leq 2(8N)^{1/\alpha}$, we have $\lfloor\bigl( \frac{(\alpha-1)N}{\alpha} \bigr)^{1/\alpha}\rfloor\geq\frac{1}{2}(\frac{\alpha-1}{\alpha})^{1/\alpha}N^{1/\alpha}\geq\frac{1}{4\cdot 8^{1/\alpha}}(\frac{\alpha-1}{\alpha})^{1/\alpha}p$; while $\frac{1}{4\cdot 8^{1/\alpha}}(\frac{\alpha-1}{\alpha})^{1/\alpha}\leq 1$. Therefore $\min\{p,\lfloor\bigl( \frac{(\alpha-1)N}{\alpha} \bigr)^{1/\alpha}\rfloor\}\geq \frac{1}{4\cdot 8^{1/\alpha}}(\frac{\alpha-1}{\alpha})^{1/\alpha}p$, hence
\begin{align*}
    \VAR^\FSD(1) \gtrsim_\alpha \log\Bigl(1+\frac{p}{N}\Bigr),\qquad \mbox{ when }\qquad p\leq \min\{N, 2(8N)^{1/\alpha}\}.
\end{align*}
Combining these three cases, for every $N>N_0$, we obtain $r^2(V_{J_\star(1)},V_{J_\star^c(1)})\gtrsim_\alpha \log(1+\frac{p}{N})$.

\paragraph{End of the proof of \texorpdfstring{\eqref{eq:laplace-decay-examples}}{Add-1}}In the three cases $p\geq N$, $2(8N)^{1/\alpha}\leq p\leq N$, and $p\leq 2(8N)^{1/\alpha}$, using respectively the lower bounds on $\Bias_{J_\star(1)}$, $\Bias_{J_\star^c(1)}$, and $\VAR^\FSD(1)$, the nonnegativity of the remaining terms, the upper bound in \cite{mourtada2026}, and Theorem~\ref{thm:main}, we obtain the claim. Here we use $\frac14\log(1+\frac pN)\leq\log\frac{N+p}{N+1}\leq\log(1+\frac pN)$ for $p\geq2$. Thus,
\begin{align}\tag{\ref{eq:laplace-decay-examples}}
\forall p\geq2,\quad\forall N>N_0,\qquad r^2(V_{J_\star(1)},V_{J_\star(1)^c})\asymp\bE\KL(\bbeta^\star\|\widehat\bbeta^1)\asymp_\alpha\log\frac{N+p}{N+1}.
\end{align}

\subsection{Proof of Corollary~\texorpdfstring{\ref{coro:Zipf}}{Zipf}}\label{sec:proof_coro_Zipf}

Here $p=\lfloor N^\iota\rfloor$. We first prove the following lemma.
\begin{Lemma}\label{lem:polynomial-rate-ranges}
    Let $\alpha>1$, $\iota>0$, $p=\lfloor N^\iota\rfloor$, and $\beta_j^\star=j^{-\alpha}(\sum_{k=1}^p k^{-\alpha})^{-1}$.
    \begin{enumerate}
        \item When $\iota\leq\frac{1}{\alpha}$. For any $0<c\leq C<\infty$, there exists $N_0=N_0(\alpha,\iota,c,C)$ such that for any $N>N_0$, and any $c\leq\lambda\leq C$,
        \begin{align*}
            r^2(V_{J_\star(\lambda)},V_{J_\star^c(\lambda)}) \asymp_{\alpha,c,C}\frac{p}{N}.
        \end{align*}
        \item When $\iota>1/\alpha$, there exists $N_0=N_0(\alpha,\iota)$ such that for any $N>N_0$ and any $0<\lambda\leq \min\{1,\frac{N}{p}\}$,
        \begin{align*}
            r^2(V_{J_\star(\lambda)},V_{J_\star^c(\lambda)}) \asymp_{\alpha,\iota}\frac{\lambda p}{N} + N^{\frac{1}{\alpha}-1}\log\Bigl(1+\frac{1}{\lambda}\Bigr).
        \end{align*}
    \end{enumerate}
\end{Lemma}
\begin{proof}
Fix
\begin{align*}
N_0=\begin{cases}\bigl\lceil\max\{2,(1+C)^{\alpha/(\alpha-1)},\frac{2^{2\alpha+2}\alpha(1+C)}{\alpha-1},(\frac{2^{\alpha+2}\alpha}{\alpha-1})^{1/\iota}\}\bigr\rceil,&0<\iota\leq\frac1\alpha,\\ \bigl\lceil\max\{2,\frac{2^{2\alpha+2}\alpha}{\alpha-1},(\frac{2^{\alpha+2}\alpha}{\alpha-1})^{1/\iota},2^{1/(\iota-1/\alpha)}\}\bigr\rceil,&\iota>\frac1\alpha,\end{cases}
\end{align*}
and suppose $N>N_0$. In both cases, $\lambda p\leq N$: this follows from the range of $\lambda$ when $\iota>1/\alpha$, and from $\lambda p\leq CN^{1/\alpha}\leq N$ when $\iota\leq1/\alpha$. The choice of $N_0$ also ensures $p\geq\frac{2^{\alpha+1}\alpha}{\alpha-1}$; when $\iota>1/\alpha$, it ensures $p\geq N^{1/\alpha}$. We use these inequalities below.
\paragraph{Calculation of \texorpdfstring{$J_\star(\lambda)$}{the estimation index set}.}
Recall that in both cases, $\lambda p\leq N$, hence
\begin{equation}\tag{\ref{eq:examples-index-set}}
    \begin{aligned}J_\star(\lambda)&=\biggl\{1\leq j\leq p:j<\biggl(\frac{4(N+\lambda p)}{\lambda\sum_{k=1}^p k^{-\alpha}}\biggr)^{1/\alpha}\biggr\},\mbox{ and } |J_\star(\lambda)|=\min\biggl\{p,\biggl\lceil\biggl(\frac{4(N+\lambda p)}{\lambda\sum_{k=1}^p k^{-\alpha}}\biggr)^{1/\alpha}\biggr\rceil-1\biggr\} \asymp_\alpha \min\biggl\{p,\Bigl(\frac{N}{\lambda}\Bigr)^{1/\alpha}\biggr\}.
    \end{aligned}
\end{equation}

\subsubsection{Calculation of \texorpdfstring{$\BIAS^\FSD$}{bias}}
\paragraph{Calculation of \texorpdfstring{$\mathrm{Bias}_{J_\star(\lambda)}$}{the estimation subspace bias}.}
Recall that
\begin{align}\tag{\ref{eq:fsd-components}}
    \mathrm{Bias}_{J_\star}=\sum_{j\in J_\star}\biggl[\beta_j^\star\log\frac{(N+\lambda p)\beta_j^\star}{N\beta_j^\star+\lambda}+\frac{N\beta_j^\star+\lambda}{N+\lambda p}-\beta_j^\star\biggr].
\end{align}Notice that $\frac{N\beta_j^\star+\lambda}{N+\lambda p} - \beta_j^\star = \frac{\lambda(1-p\beta_j^\star)}{N+\lambda p}$. We first show that for any $j\in J_\star(\lambda)$,
\begin{align}\label{eq:identity_Bias_j}
    &\beta_j^\star\log\frac{(N+\lambda p)\beta_j^\star}{N\beta_j^\star+\lambda}+\frac{N\beta_j^\star+\lambda}{N+\lambda p}-\beta_j^\star = \frac{\lambda^2(1-p\beta_j^\star)^2}{(N+\lambda p)^2}\int_0^1\frac{t\mathrm{d}t}{(1-t)\beta_j^\star + t\frac{N\beta_j^\star+\lambda}{N+\lambda p}}.
\end{align}Indeed, for any $0\leq t\leq 1$,
\begin{align*}
    &\frac{\mathrm{d}}{\mathrm{d}t}\biggl( \beta_j^\star\log\frac{\beta_j^\star}{ (1-t)\beta_j^\star + t\frac{N\beta_j^\star+\lambda}{N+\lambda p} } + t\Bigl( \frac{N\beta_j^\star+\lambda}{N+\lambda p} - \beta_j^\star \Bigr) \biggr)=-\frac{\beta_j^\star\Bigl(\frac{N\beta_j^\star+\lambda}{N+\lambda p} - \beta_j^\star\Bigr)}{(1-t)\beta_j^\star + t\frac{N\beta_j^\star+\lambda}{N+\lambda p}} + \frac{N\beta_j^\star+\lambda}{N+\lambda p} - \beta_j^\star\\
    &= \Bigl(\frac{N\beta_j^\star+\lambda}{N+\lambda p} - \beta_j^\star\Bigr)\frac{(1-t)\beta_j^\star + t\frac{N\beta_j^\star+\lambda}{N+\lambda p}-\beta_j^\star}{(1-t)\beta_j^\star + t\frac{N\beta_j^\star+\lambda}{N+\lambda p}} =\frac{t\Bigl(\frac{N\beta_j^\star+\lambda}{N+\lambda p} - \beta_j^\star\Bigr)^2}{(1-t)\beta_j^\star + t\frac{N\beta_j^\star+\lambda}{N+\lambda p}} =\frac{t\lambda^2(1-p\beta_j^\star)^2}{(N+\lambda p)^2\bigl((1-t)\beta_j^\star+t\frac{N\beta_j^\star+\lambda}{N+\lambda p}\bigr)}.
\end{align*}Hence \eqref{eq:identity_Bias_j} follows from the fundamental theorem of calculus, together with the fact that when $t=0$, $ \beta_j^\star\log\frac{\beta_j^\star}{ (1-t)\beta_j^\star + t\frac{N\beta_j^\star+\lambda}{N+\lambda p} } + t\Bigl( \frac{N\beta_j^\star+\lambda}{N+\lambda p} - \beta_j^\star \Bigr)=0$, while when $t=1$, it equals $\beta_j^\star\log\frac{(N+\lambda p)\beta_j^\star}{N\beta_j^\star+\lambda}+\frac{N\beta_j^\star+\lambda}{N+\lambda p}-\beta_j^\star$.

Now, we estimate the integrand. Recall that 
\begin{align}\tag{\ref{eq:decomposition}}
    \begin{aligned}
        &J_\star(\lambda)=\bigg\{1\leq j\leq p:\beta_j^\star>\frac\lambda{4(N+\lambda p)}\bigg\},
    \end{aligned}
\end{align}we then know that for any $j\in J_\star(\lambda),$ $\beta_j^\star>\frac{\lambda}{4(N+\lambda p)}$; hence
$\frac{N\beta_j^\star+\lambda}{N+\lambda p}\leq \beta_j^\star+\frac{\lambda}{N+\lambda p}<5\beta_j^\star$. Combining with $\frac{N\beta_j^\star+\lambda}{N+\lambda p}\geq \frac{N\beta_j^\star}{N+\lambda p}\geq \frac{1}{2}\beta_j^\star$, there holds
\begin{align*}
    \forall 0\leq t\leq 1,\qquad \frac{1}{2}\beta_j^\star\leq (1-\frac{1}{2}t)\beta_j^\star = (1-t)\beta_j^\star + \frac{t}{2}\beta_j^\star\leq (1-t)\beta_j^\star + t\frac{N\beta_j^\star+\lambda}{N+\lambda p} \leq 5\beta_j^\star.
\end{align*}Combining with $\int_0^1t\mathrm{d}t=\frac{1}{2}$,
\begin{align*}
    &\frac{1}{10\beta_j^\star} \leq \int_0^1\frac{t\mathrm{d}t}{(1-t)\beta_j^\star + t\frac{N\beta_j^\star+\lambda}{N+\lambda p}} \leq \frac{1}{\beta_j^\star}.
\end{align*}Since $\lambda p\leq N$, \eqref{eq:identity_Bias_j} implies that
\begin{align}\label{eq:equivalent_Bias_J_star}
    \frac{\lambda^2}{40N^2}\sum_{j\in J_\star(\lambda)}\frac{(1-p\beta_j^\star)^2}{\beta_j^\star} \leq \Bias_{J_\star(\lambda)}\leq \frac{\lambda^2}{N^2}\sum_{j\in J_\star(\lambda)}\frac{(1-p\beta_j^\star)^2}{\beta_j^\star}.
\end{align}Now since $(1-p\beta_j^\star)^2\leq 2+2p^2(\beta_j^\star)^2$; $\sum_{j\in J_\star(\lambda)}\frac{1}{\beta_j^\star} = (\sum_{k=1}^pk^{-\alpha})\sum_{j=1}^{|J_\star(\lambda)|}j^\alpha\leq \frac{\alpha}{\alpha-1}|J_\star(\lambda)|^{\alpha+1}$, we obtain from \eqref{eq:equivalent_Bias_J_star} that
\begin{align}\label{eq:upper_Bias_J_star}
    \Bias_{J_\star(\lambda)}\leq\frac{2\alpha}{\alpha-1}\frac{\lambda^2}{N^2}\Bigl(p^2 + |J_\star(\lambda)|^{\alpha+1}\Bigr).
\end{align}For the lower bound, we use $1\in J_\star(\lambda)$; $\beta_1^\star\geq\frac{\alpha-1}{\alpha}$, hence $p\beta_1^\star\geq 2$ by the choice of $N_0$, and hence $(1-p\beta_1^\star)^2=(p\beta_1^\star-1)^2\geq (p\beta_1^\star/2)^2$, and
\begin{align*}
    \sum_{j\in J_\star(\lambda)}\frac{(1-p\beta_j^\star)^2}{\beta_j^\star}\geq\frac{(1-p\beta_1^\star)^2}{\beta_1^\star}\geq \frac{p^2\beta_1^\star}{4}\geq \frac{\alpha-1}{4\alpha}p^2.
\end{align*}When $2(2p)^{1/\alpha}> |J_\star(\lambda)|$, combining \eqref{eq:equivalent_Bias_J_star}, \eqref{eq:upper_Bias_J_star}, and the above estimation, we obtain $\Bias_{J_\star(\lambda)}\asymp_\alpha\frac{\lambda^2}{N^2}(p^2+|J_\star(\lambda)|^{\alpha+1})$; otherwise, when $2(2p)^{1/\alpha}\le |J_\star(\lambda)|$, for any $\frac{1}{2}|J_\star(\lambda)|\leq j\leq  |J_\star(\lambda)|$, using $\beta_j^\star\leq j^{-\alpha}$, we obtain $p\beta_j^\star\leq pj^{-\alpha}\leq p(\frac{2}{|J_\star(\lambda)|})^\alpha\leq \frac{1}{2}$. Hence $(1-p\beta_j^\star)^2\geq\frac{1}{4}$, and
\begin{align*}
    \sum_{j\in J_\star(\lambda)}\frac{(1-p\beta_j^\star)^2}{\beta_j^\star} \geq 2^{-(\alpha+3)}|J_\star(\lambda)|^{\alpha+1} \gtrsim_\alpha p^2 + |J_\star(\lambda)|^{\alpha+1}.
\end{align*}In summary,
\begin{equation}\label{eq:examples-tuned-estimation-bias}
    \begin{aligned}\frac{\lambda^2}{40N^2}\sum_{j\in J_\star(\lambda)}\frac{(1-p\beta_j^\star)^2}{\beta_j^\star}&\leq\mathrm{Bias}_{J_\star(\lambda)}\leq\frac{\lambda^2}{N^2}\sum_{j\in J_\star(\lambda)}\frac{(1-p\beta_j^\star)^2}{\beta_j^\star},\mbox{ and }\mathrm{Bias}_{J_\star(\lambda)}\asymp_\alpha\frac{\lambda^2}{N^2}\bigl(p^2+|J_\star(\lambda)|^{\alpha+1}\bigr).\end{aligned}
\end{equation}

\paragraph{Calculation of \texorpdfstring{$\mathrm{Bias}_{J_\star(\lambda)^c}$}{the free subspace bias}.}We establish an upper bound for every $\lambda>0$ and a lower bound when $|J_\star(\lambda)|\leq \frac{1}{2}p$. Recall that
\begin{align}\tag{\ref{eq:fsd-components}}
    \mathrm{Bias}_{J_\star^c(\lambda)}=\frac{\lambda|J_\star^c(\lambda)|}{N+\lambda p}.
\end{align}
By definition, $\Bias_{J_\star(\lambda)^c} \leq \frac{\lambda p}{N}.$ For the lower bound, from $|J_\star(\lambda)|\leq \frac{1}{2}p$ and $\lambda p\leq N$,
\begin{align*}
    \Bias_{J_\star^c(\lambda)} = \frac{\lambda(p-|J_\star(\lambda)|)}{N+\lambda p} \geq \frac{\lambda p}{4N}.
\end{align*}In summary,
\begin{equation}\label{eq:examples-tuned-free-bias}
    \mathrm{Bias}_{J_\star(\lambda)^c}\leq\frac{\lambda p}N,\quad\mbox{ and }\quad  \frac{\lambda p}{4N}\leq\mathrm{Bias}_{J_\star(\lambda)^c}\leq\frac{\lambda p}N\quad\mbox{if }|J_\star(\lambda)|\leq\frac p2.
\end{equation}
\subsubsection{Calculation of \texorpdfstring{$\mathrm{VAR}^{\mathrm{FSD}}(\lambda)$}{the FSD variance}.}
We recall that
\begin{align}\tag{\ref{eq:fsd-components}}
    \mathrm{VAR}^{\mathrm{FSD}}=\sum_{j\in J_\star}\biggl[\frac{N(\beta_j^\star)^2(1-\beta_j^\star)}{(N\beta_j^\star+\lambda)(N\beta_j^\star+\lambda+1)}+\mathbf1_{\{\lambda<1\}}\beta_j^\star(1-\beta_j^\star)^N\log\biggl(1+\frac{N\beta_j^\star}{\lambda}\biggr)\biggr].
\end{align}
We consider two cases.
\paragraph{When \texorpdfstring{$\iota>\frac{1}{\alpha}$}{iota large}, and \texorpdfstring{$0<\lambda\leq 1$}{lambda small}.}
\begin{enumerate}
    \item Lower bound. 
    Take
    \begin{align*}
        I=\Bigl\{ j\in [p]:\, \frac{1}{2}\Bigl(\frac{(\alpha-1)N}{2\alpha}\Bigr)^{\frac{1}{\alpha}}\leq j \leq \Bigl(\frac{(\alpha-1)N}{2\alpha}\Bigr)^{\frac{1}{\alpha}} \Bigr\}.
    \end{align*}Then $|I|\geq\frac{1}{4}(\frac{\alpha-1}{2\alpha})^{\frac{1}{\alpha}}N^{\frac{1}{\alpha}}$. Since $\frac{\alpha-1}{\alpha}j^{-\alpha}\leq \beta_j^\star\leq j^{-\alpha}$, we know that for any $j\in I$, $2\leq N\beta_j^\star\leq \frac{2^{\alpha+1}\alpha}{\alpha-1}$. Since $0<\lambda\leq1$, $\frac{\lambda}{4(N+\lambda p)}\leq\frac{1}{4N}$. Therefore, $I\subset J_\star(\lambda)$, where we recall that 
\begin{align}\tag{\ref{eq:decomposition}}
    \begin{aligned}
        &J_\star(\lambda)=\bigg\{1\leq j\leq p:\beta_j^\star>\frac\lambda{4(N+\lambda p)}\bigg\},
    \end{aligned}
\end{align}The choice of $N_0$ also gives
    \begin{align*}
        \forall j\in I,\qquad N\beta_j^\star\geq 2,\quad N\beta_j^\star+\lambda\leq\frac{3}{2}N\beta_j^\star,\quad N\beta_j^\star+\lambda+1\leq 2N\beta_j^\star,\quad\mbox{ and }\quad \beta_j^\star \leq \frac{2^{\alpha+1}\alpha}{\alpha-1}\frac{1}{N}\leq\frac{1}{2},
    \end{align*}
    When $0<\lambda<1$, from $\log(1-u)\geq -2u$ for any $0\leq u\leq \frac{1}{2}$, we have $(1-\beta_j^\star)^N\geq\exp(-2N\beta_j^\star)\gtrsim_\alpha1$. Hence
    \begin{align*}
        &\sum_{j\in J_\star(\lambda)} \beta_j^\star(1-\beta_j^\star)^N\log\biggl(1+\frac{N\beta_j^\star}{\lambda}\biggr)\gtrsim_\alpha \sum_{j\in I}\frac{1}{N}\log(2)\log\Bigl(1+\frac{1}{\lambda}\Bigr)\gtrsim_\alpha N^{\frac{1}{\alpha}-1}\log\Bigl(1+\frac{1}{\lambda}\Bigr). 
    \end{align*}
    Next consider $\lambda=1$. We obtain that
    \begin{align*}
        \sum_{j\in J_\star(\lambda)}\frac{N(\beta_j^\star)^2(1-\beta_j^\star)}{(N\beta_j^\star+\lambda)(N\beta_j^\star+\lambda+1)} \gtrsim\sum_{j\in I} \frac{1}{N} \gtrsim_\alpha N^{\frac{1}{\alpha}-1}\log\Bigl(1+\frac{1}{\lambda}\Bigr),
    \end{align*}where we used the fact that $\lambda=1$. In summary,
    \begin{align}\label{eq:lower_bound_iota_large}
        \forall \iota>\frac{1}{\alpha},\quad \forall 0<\lambda\leq 1,\qquad \VAR^\FSD(\lambda)\gtrsim_\alpha N^{1/\alpha-1}\log(1+\frac{1}{\lambda}).
    \end{align}

    \item Upper bound. We first bound
    \begin{align*}
        \sum_{j\in J_\star(\lambda)}\frac{N(\beta_j^\star)^2(1-\beta_j^\star)}{(N\beta_j^\star+\lambda)(N\beta_j^\star+\lambda+1)}
    \end{align*}from above. Using the fact that for any $j\in J_\star(\lambda)$, $1-\beta_j^\star\leq 1$, $(N\beta_j^\star+\lambda)(N\beta_j^\star+\lambda+1)\geq (N\beta_j^\star)^2$ (and is also larger than $N\beta_j^\star$), we obtain that
    \begin{align*}
        \sum_{j\in J_\star(\lambda)}\frac{N(\beta_j^\star)^2(1-\beta_j^\star)}{(N\beta_j^\star+\lambda)(N\beta_j^\star+\lambda+1)} &\leq \sum_{j\in J_\star(\lambda)}\min\Bigl\{\frac{1}{N},\beta_j^\star\Bigr\}\leq \sum_{j\in J_\star(\lambda), j\leq \lceil N^{1/\alpha}\rceil}\frac{1}{N} + \sum_{j\in J_\star(\lambda),j>\lceil N^{1/\alpha}\rceil}\beta_j^\star\\
        &\leq 2N^{\frac{1}{\alpha}-1} + \sum_{j>\lceil N^{1/\alpha}\rceil} j^{-\alpha}\lesssim_\alpha N^{\frac{1}{\alpha}-1}.
    \end{align*}Now we separate the sum
    \begin{align*}
        \sum_{j\in J_\star}\beta_j^\star(1-\beta_j^\star)^N\log\biggl(1+\frac{N\beta_j^\star}{\lambda}\biggr)
    \end{align*}into two parts.
    \begin{enumerate}
        \item When $j\leq\lfloor N^{1/\alpha}\rfloor$. For any $k=0,1,\cdots$, let
        \begin{align*}
            I_k = \Bigl\{ j\leq \lfloor N^{\frac{1}{\alpha}}\rfloor:\, 2^{-k-1}N^{\frac{1}{\alpha}} < j \leq 2^{-k}N^{\frac{1}{\alpha}} \Bigr\}.
        \end{align*}Then $|I_k|\leq 2^{1-k}N^{1/\alpha}$. Moreover, on each $I_k$, $\frac{\alpha-1}{\alpha}2^{\alpha k}N^{-1} \leq \beta_j^\star < 2^{\alpha(k+1)}N^{-1}$, hence
        \begin{align*}
            (1-\beta_j^\star)^N\leq \exp(-N\beta_j^\star)\leq \exp(-\frac{\alpha-1}{\alpha}2^{k\alpha}),\quad\mbox{ and }\quad \log\Bigl(1+\frac{N\beta_j^\star}{\lambda}\Bigr) \leq(1+(k+1)\alpha)\log\Bigl(1+\frac1\lambda\Bigr). 
        \end{align*}As a result,
        \begin{align*}
            \sum_{\substack{j\in J_\star\\ j\leq \lfloor N^{1/\alpha}\rfloor}}\beta_j^\star(1-\beta_j^\star)^N\log\biggl(1+\frac{N\beta_j^\star}{\lambda}\biggr) &\leq2^{\alpha+1}N^{\frac{1}{\alpha}-1}\log\Bigl(1+\frac{1}{\lambda}\Bigr)\sum_{k=0}^\infty 2^{k(\alpha-1)}(1+(k+1)\alpha)\exp\Bigl(-\frac{\alpha-1}{\alpha}2^{k\alpha}\Bigr)\\
            &\lesssim_\alpha N^{\frac{1}{\alpha}-1}\log\Bigl(1+\frac{1}{\lambda}\Bigr),
        \end{align*}The series converges because $\alpha>1$.
        \item When $j>\lfloor N^{1/\alpha}\rfloor$. Since $\beta_j^\star\leq j^{-\alpha}$, we obtain $\log(1+N\beta_j^\star)\leq 1$. Together with $\log(1+\frac{N\beta_j^\star}{\lambda})\leq \log(\frac{1}{\lambda})+\log(1+N\beta_j^\star)$, we obtain
        \begin{align*}
            \sum_{\substack{j\in J_\star\\ j> \lfloor N^{1/\alpha}\rfloor}}\beta_j^\star(1-\beta_j^\star)^N\log\biggl(1+\frac{N\beta_j^\star}{\lambda}\biggr) &\leq \sum_{j>\lfloor N^{1/\alpha}\rfloor} j^{-\alpha}\cdot 1^N \Bigl( 1+\log\bigl(\frac{1}{\lambda}\bigr) \Bigr).
        \end{align*}
    \end{enumerate}
    Combining the above, we obtain
\begin{align}\label{eq:upper_bound_iota_large}
        \forall \iota>\frac{1}{\alpha},\quad \forall 0<\lambda\leq 1,\qquad \VAR^\FSD(\lambda)\asymp_\alpha N^{1/\alpha-1}\log(1+\frac{1}{\lambda}).
    \end{align}
\end{enumerate}

\paragraph{When \texorpdfstring{$\iota\leq \frac{1}{\alpha}$}{iota small}, and \texorpdfstring{$c\leq\lambda\leq C$}{lambda large}.}
In this case, our objective is to show that $\VAR^\FSD(\lambda)\asymp_{\alpha,c,C}\frac{p}{N}$. From $1-\beta_j^\star\leq 1$ and $(N\beta_j^\star+\lambda)(N\beta_j^\star+\lambda+1)\geq (N\beta_j^\star)^2$, there holds
\begin{align*}
    \sum_{j\in J_\star(\lambda)}\frac{N(\beta_j^\star)^2(1-\beta_j^\star)}{(N\beta_j^\star+\lambda)(N\beta_j^\star+\lambda+1)}\leq \frac{|J_\star(\lambda)|}{N}\leq \frac{p}{N}.
\end{align*}When $c\leq\lambda<1$, by $\log(1+x)\leq x$,
\begin{align*}
    \frac{\beta_j^\star(1-\beta_j^\star)^N\log(1+\frac{N\beta_j^\star}{\lambda})}{\frac{N(\beta_j^\star)^2(1-\beta_j^\star)}{(N\beta_j^\star+\lambda)(N\beta_j^\star+\lambda+1)}} \leq \frac{(N\beta_j^\star+\lambda)(N\beta_j^\star+\lambda+1)}{\lambda}(1-\beta_j^\star)^{N-1} \lesssim_{\alpha,c,C}1,
\end{align*}where we have used $(1-\beta_j^\star)^{N-1}\leq \exp(1-N\beta_j^\star)$. Therefore, $\sum_{j\in J_\star}\Bigl[\mathbf1_{\{\lambda<1\}}\beta_j^\star(1-\beta_j^\star)^N\log\Bigl(1+\frac{N\beta_j^\star}{\lambda}\Bigr)\Bigr]$ is bounded by a constant multiple of $\sum_{j\in J_\star}\Bigl[\frac{N(\beta_j^\star)^2(1-\beta_j^\star)}{(N\beta_j^\star+\lambda)(N\beta_j^\star+\lambda+1)}\Bigr]$. When $C\geq\lambda\geq1$, this term vanishes. In summary,
\begin{align}\label{eq:upper_variance_iota_small}
    &\forall \iota\leq\frac1\alpha,\quad \forall c\leq\lambda\leq C,\qquad \VAR^\FSD(\lambda)\lesssim_{\alpha,c,C}\frac{p}{N}.
\end{align}

For the lower bound, we only consider $\sum_{j\in J_\star}\Bigl[\frac{N(\beta_j^\star)^2(1-\beta_j^\star)}{(N\beta_j^\star+\lambda)(N\beta_j^\star+\lambda+1)}\Bigr]$, since $\sum_{j\in J_\star}\Bigl[\mathbf1_{\{\lambda<1\}}\beta_j^\star(1-\beta_j^\star)^N\log\Bigl(1+\frac{N\beta_j^\star}{\lambda}\Bigr)\Bigr]$ is non-negative. Let
\begin{align*}
    I'=\Big\{ j\in[p]:\, 1\leq j\leq \Bigl\lfloor \Bigl( \frac{(\alpha-1)N}{\alpha(1+C)}\Bigr)^{1/\alpha} \Bigr\rfloor \Big\}.
\end{align*}Recall $\lambda p\leq N$, $\sum_{k=1}^pk^{-\alpha}\leq\frac{\alpha}{\alpha-1}$, then for any $j\in I'$, $\beta_j^\star=\frac{j^{-\alpha}}{\sum_{k=1}^p k^{-\alpha}}\geq \frac{\alpha-1}{\alpha}j^{-\alpha}\geq \frac{1+C}{N}$. On the other hand, from $\lambda\leq C$ we know that $\frac{\lambda}{4(N+\lambda p)}\leq\frac{C}{4N}<\frac{1+C}{N}\leq \beta_j^\star$. Combining with
\begin{align}\tag{\ref{eq:examples-index-set}}
    J_\star(\lambda)&=\biggl\{1\leq j\leq p:j<\biggl(\frac{4(N+\lambda p)}{\lambda\sum_{k=1}^p k^{-\alpha}}\biggr)^{1/\alpha}\biggr\},
\end{align}we know that $I'\subset J_\star(\lambda)$.
Moreover, for any $j\in I'$,  $j^\alpha\leq \frac{\alpha-1}{\alpha(1+C)}N$, hence $N\beta_j^\star=Nj^{-\alpha}(\sum_{k=1}^pk^{-\alpha})^{-1}\geq\frac{\alpha-1}{\alpha}Nj^{-\alpha}\geq 1+C$. Combining with $1-\beta_j^\star\geq 1-\beta_1^\star \geq 1-\frac{1}{1+2^{-\alpha}} \geq  2^{-\alpha-1}$, and for any $j\in I'$, $\beta_j^\star\geq\frac{\alpha-1}{\alpha}j^{-\alpha}$,
\begin{align*}
    \forall j\in I',\qquad \frac{N(\beta_j^\star)^2(1-\beta_j^\star)}{(N\beta_j^\star+\lambda)(N\beta_j^\star+\lambda+1)}\gtrsim_\alpha \frac{1}{N},
\end{align*}where we have used $\lambda\leq C$ and $N\beta_j^\star\geq 1+C$ to derive $N\beta_j^\star+\lambda\leq 2N\beta_j^\star$, and $N\beta_j^\star+\lambda+1\leq 2N\beta_j^\star$.
Since $p\leq N^\iota\leq N^{1/\alpha}$, $|I'|\gtrsim_{\alpha,C}p$, hence
\begin{align}\label{eq:lower_variance_iota_small}
    &\forall \iota\leq\frac1\alpha,\quad \forall c\leq\lambda\leq C,\qquad \VAR^\FSD(\lambda)\gtrsim_\alpha |I'|\frac{1}{N} \asymp_{\alpha,C}\frac{p}{N}.
\end{align}

\subsubsection{End of the proof of Lemma~\texorpdfstring{\ref{lem:polynomial-rate-ranges}}{4}}
Recall that
\begin{equation}\tag{\ref{eq:examples-tuned-estimation-bias}}
    \begin{aligned}
    \mathrm{Bias}_{J_\star(\lambda)}\asymp_\alpha\frac{\lambda^2}{N^2}\bigl(p^2+|J_\star(\lambda)|^{\alpha+1}\bigr),\end{aligned}
\end{equation}
\begin{equation}\tag{\ref{eq:examples-tuned-free-bias}}
    \mathrm{Bias}_{J_\star(\lambda)^c}\leq\frac{\lambda p}N,\quad\mbox{ and }\quad  \frac{\lambda p}{4N}\leq\mathrm{Bias}_{J_\star(\lambda)^c}\leq\frac{\lambda p}N\quad\mbox{if }|J_\star(\lambda)|\leq\frac p2,
\end{equation}
\begin{align}\tag{\ref{eq:upper_bound_iota_large}}
        \forall \iota>\frac{1}{\alpha},\quad \forall 0<\lambda\leq 1,\qquad \VAR^\FSD(\lambda)\asymp_\alpha N^{1/\alpha-1}\log(1+\frac{1}{\lambda}),
\end{align}and
\begin{align}\tag{\ref{eq:lower_variance_iota_small}}
    &\forall \iota\leq\frac1\alpha,\quad \forall c\leq\lambda\leq C,\qquad \VAR^\FSD(\lambda)\gtrsim_\alpha |I'|\frac{1}{N} \asymp_{\alpha,C}\frac{p}{N}.
\end{align}
\begin{enumerate}
    \item When $\iota\leq\frac{1}{\alpha}$, and $c\leq \lambda\leq C$. We obtain
    \begin{align*}
        \Bias_{J_\star(\lambda)}\lesssim_{\alpha,C}\frac{p}{N},\quad \Bias_{J_\star^c(\lambda)}\lesssim_C \frac{p}{N},\quad\mbox{ and }\quad \VAR^\FSD(\lambda)\asymp_{\alpha,c,C}\frac{p}{N}.
    \end{align*}
    \item When $\iota>\frac{1}{\alpha}$, and $0<\lambda\leq\min\{1,\frac{N}{p}\}$. We obtain
    \begin{align*}
        \Bias_{J_\star(\lambda)} \lesssim_\alpha \frac{\lambda p}{N}+N^{\frac{1}{\alpha}-1}\log\Bigl(1+\frac{1}{\lambda}\Bigr),\quad \Bias_{J_\star^c(\lambda)}\leq\frac{\lambda p}{N},\mbox{ and } \VAR^\FSD(\lambda)\asymp_\alpha   N^{\frac{1}{\alpha}-1}\log\Bigl(1+\frac{1}{\lambda}\Bigr).
    \end{align*}To finish the proof of Lemma~\ref{lem:polynomial-rate-ranges}, notice that when $|J_\star(\lambda)|\leq \frac{1}{2}p$, \eqref{eq:examples-tuned-free-bias} together with \eqref{eq:upper_bound_iota_large} yield $r^2(V_{J_\star(\lambda)},V_{J_\star^c(\lambda)})\gtrsim_\alpha \frac{\lambda p}{N}+N^{1/\alpha-1}\log(1+\frac{1}{\lambda})$; while when $|J_\star(\lambda)|> \frac{1}{2}p$, by the fact that $j\mapsto \beta_j^\star$ is decreasing, we obtain that $\lceil p/2\rceil \in J_\star(\lambda)$. By the definition of $J_\star(\lambda)$,
    \begin{align*}
        \frac{\lambda}{8N} \leq \frac{\lambda}{4(N+\lambda p)} < \beta_{\lceil p/2\rceil}^\star \leq 2^\alpha p^{-\alpha}.
    \end{align*}Therefore, from $p\geq N^{1/\alpha}$,  $\lambda\leq 1$, and $\log(1+\frac{1}{\lambda})\geq\log(2)$, we obtain that
    \begin{align*}
        \frac{\lambda p}{N} \leq 2^{\alpha+3}p^{1-\alpha}\leq 2^{\alpha+3}N^{1/\alpha-1}\lesssim_\alpha N^{1/\alpha-1}\log\Bigl(1+\frac{1}{\lambda}\Bigr).
    \end{align*}Therefore, \eqref{eq:upper_bound_iota_large} implies that $r^2(V_{J_\star(\lambda)},V_{J_\star^c(\lambda)})\gtrsim_\alpha \frac{\lambda p}{N}+N^{1/\alpha-1}\log(1+\frac{1}{\lambda})$.
\end{enumerate}
\end{proof}

\subsubsection{Calculation of the oracle parameter}

\begin{Lemma}\label{lem:polynomial-oracle-optimality}
    Let $\alpha>1$, $\iota>0$, $p=\lfloor N^\iota\rfloor$, and $\beta_j^\star=j^{-\alpha}(\sum_{k=1}^p k^{-\alpha})^{-1}$. Set $\lambda^\star=\min\{1,N^{1/\alpha}/p\}$. There exists $N_0=N_0(\alpha,\iota)$ such that, for every $N>N_0$,
    \begin{align*}
        r^2(V_{J_\star(\lambda^\star)},V_{J_\star(\lambda^\star)^c})\asymp_{\alpha,\iota}\inf\bigl\{r^2(V_{J_\star(\lambda)},V_{J_\star(\lambda)^c}):\lambda>0\bigr\}.
    \end{align*}
\end{Lemma}
\begin{proof}
    Fix
    \begin{align*}
    N_0=\begin{cases}\bigl\lceil\max\{2,2^{\alpha/(\alpha-1)},\frac{2^{2\alpha+3}\alpha}{\alpha-1},(\frac{2^{\alpha+2}\alpha}{\alpha-1})^{1/\iota},2^{2\alpha/(\iota(\alpha-1))}\}\bigr\rceil,&0<\iota\leq\frac1\alpha,\\ \bigl\lceil\max\{2,\frac{2^{2\alpha+2}\alpha}{\alpha-1},(\frac{2^{\alpha+2}\alpha}{\alpha-1})^{1/\iota},4^{1/(\iota-1/\alpha)},2^{(2\alpha+3)/((\iota-1/\alpha)(\alpha-1/2))},2^{(2\alpha+2)/(\iota(\alpha-1))},(1+\frac{2\alpha\iota}{\alpha-1})^{2\alpha/(\alpha-1)}\}\bigr\rceil,&\iota>\frac1\alpha,\end{cases}
    \end{align*}
    and suppose $N>N_0$. We separate the proof into two cases.
    \begin{enumerate}
        \item When $\iota\leq\frac{1}{\alpha}$. Since $p=\lfloor N^\iota\rfloor$, we know that $p\leq N^{1/\alpha}$. In this case, $\lambda^\star=1$, and $r^2(V_{J_\star(\lambda^\star)},V_{J_\star^c(\lambda^\star)})\asymp_\alpha \frac{p}{N}$. It remains to prove that, for every $\lambda>0$, $r^2(V_{J_\star(\lambda)},V_{J_\star^c(\lambda)})\gtrsim_\alpha \frac{p}{N}$. We consider two further cases.
        \begin{enumerate}
            \item When $0<\lambda\leq N\beta_p^\star$. By $\beta_j^\star\geq\beta_p^\star$ for any $1\leq j\leq p$, we obtain that for any $1\leq j\leq p$, $\beta_j^\star\geq\beta_p^\star\geq\frac{\lambda}{N}>\frac{\lambda}{4(N+\lambda p)}$, indicating that $J_\star(\lambda)=[p]$. Moreover, by $p\leq N^{1/\alpha}$, we know that $N\beta_j^\star\geq N\beta_p^\star\geq\frac{\alpha-1}{\alpha}$. Together with $\lambda\leq N\beta_j^\star$, we know that $N\beta_j^\star+\lambda\leq 2N\beta_j^\star$ and $N\beta_j^\star+\lambda+1\leq (2+\frac{\alpha}{\alpha-1})N\beta_j^\star$; while for any $1\leq j\leq p$, $1-\beta_j^\star\geq 1-\beta_1^\star\geq \frac{2^{-\alpha}}{1+2^{-\alpha}}$. As a result,
            \begin{align*}
                \forall 0<\lambda\leq N\beta_p^\star,\qquad r^2(V_{J_\star(\lambda)},V_{J_\star^c(\lambda)})\geq \VAR^\FSD\geq \sum_{j=1}^p\frac{N(\beta_j^\star)^2(1-\beta_j^\star)}{(N\beta_j^\star+\lambda)(N\beta_j^\star+\lambda+1)}\gtrsim_\alpha \frac{p}{N}.
            \end{align*}

            \item When $\lambda\geq N\beta_p^\star$. By Proposition~\ref{prop:monotonicity}, $\lambda\in(0,\infty)\mapsto\BIAS^\FSD(\lambda)$ is nondecreasing, hence we only need to prove that $\BIAS^\FSD(N\beta_p^\star)\gtrsim_\alpha\frac{p}{N}$. Let $\underline{\lambda}=N\beta_p^\star$. Since $\beta_j^\star\geq\beta_p^\star = \frac{\underline{\lambda}}{N}>\frac{\underline{\lambda}}{4(N+\underline{\lambda}p)}$, we obtain $J_\star(\underline{\lambda})=[p]$. Moreover, from $\sum_{j=1}^p\beta_j^\star=1$, we obtain $\underline{\lambda}p=Np\beta_p^\star\leq N\sum_{j=1}^p\beta_j^\star=N$. In particular, by \eqref{eq:equivalent_Bias_J_star}, we have
            \begin{align*}
                &\BIAS^\FSD(\underline{\lambda})=\Bias_{J_\star(\underline{\lambda})} \geq \frac{\underline{\lambda}^2}{40N^2}\sum_{j=1}^p\frac{(1-p\beta_j^\star)^2}{\beta_j^\star}\geq \frac{\underline{\lambda}^2}{40N^2}\sum_{j=\lceil p/2\rceil }^p\frac{(1-p\beta_j^\star)^2}{\beta_j^\star}.
            \end{align*}By the choice of $N_0$, $p\beta_{\lceil p/2\rceil}^\star\leq 2^\alpha p^{1-\alpha}\leq\frac{1}{2}$. Together with $(\beta_j^\star)^{-1}\geq j^\alpha$, and $\beta_p^\star\geq\frac{\alpha-1}{\alpha}p^{-\alpha}$, we obtain $\BIAS^\FSD(\underline{\lambda})\gtrsim_\alpha p^{1-\alpha}\geq\frac{p}{N}$, and by Proposition~\ref{prop:monotonicity}, this implies
            \begin{align*}
                \forall \lambda\geq N\beta_p^\star,\qquad r^2(V_{J_\star(\lambda)},V_{J_\star^c(\lambda)})\gtrsim_\alpha \frac{p}{N}. 
            \end{align*}
        \end{enumerate}
        
        \item When $\iota>\frac{1}{\alpha}$. By the choice of $N_0$, we know $p = \lfloor N^\iota \rfloor > N^{1/\alpha}$, hence $\lambda^\star = \frac{N^{1/\alpha}}{p}<1$; and $\lambda^\star p = N^{1/\alpha}<N$. Therefore, $r^2(V_{J_\star(\lambda^\star)},V_{J_\star^c(\lambda^\star)})\asymp_\alpha N^{1/\alpha-1}\log(\frac{p}{N^{1/\alpha}})$. We consider two cases.
        \begin{enumerate}
            \item When $0<\lambda\leq\sqrt{\lambda^\star}$. Since $\lambda^\star<1$, by \eqref{eq:upper_bound_iota_large}, we have
            \begin{align*}
                \forall 0<\lambda\leq\sqrt{\lambda^\star},\qquad r^2(V_{J_\star(\lambda)},V_{J_\star^c(\lambda)})&\geq \VAR^\FSD(\lambda)\gtrsim_\alpha N^{1/\alpha-1}\log\Bigl(1+\frac{1}{\lambda}\Bigr) \geq N^{1/\alpha-1}\log\Bigl(1+\frac{1}{\sqrt{\lambda^\star}}\Bigr)\\
                &=N^{1/\alpha-1}\log\Bigl(1+\sqrt{\frac{p}{N^{1/\alpha}}}\Bigr) \geq \frac{1}{2}N^{1/\alpha-1}\log\Bigl(\frac{p}{N^{1/\alpha}}\Bigr).
            \end{align*}

            \item When $\lambda\geq\sqrt{\lambda^\star}$. We first use the choice of $N_0$ to show that $|J_\star(\sqrt{\lambda^\star})|\leq\frac{1}{2}p$. Notice that by \eqref{eq:examples-index-set}, for any $\lambda>0$, $|J_\star(\lambda)|\leq\Bigl( \frac{4(N+\lambda p)}{\lambda\sum_{k=1}^pk^{-\alpha}} \Bigr)^{1/\alpha}$. Since $\sum_{k=1}^pk^{-\alpha}\geq1$, we have
            \begin{align*}
                \frac{|J_\star(\sqrt{\lambda^\star})|}{p} \leq 4^{1/\alpha}\biggl(\frac{\frac{N}{\sqrt{\lambda^\star}}+p}{p^\alpha}\biggr)^{1/\alpha} = 4^{1/\alpha}\biggl( \Bigl(\frac{N^{1/\alpha}}{p}\Bigr)^{\alpha-\frac{1}{2}} + p^{1-\alpha}\biggr)^{1/\alpha}\leq\frac12.
            \end{align*}Here $p\geq N^\iota/2$, and the choice of $N_0$ bounds each of $(N^{1/\alpha}/p)^{\alpha-1/2}$ and $p^{1-\alpha}$ by $2^{-\alpha-3}$. Consequently,
            \begin{align*}
                \forall\lambda\geq\sqrt{\lambda^\star},\qquad &r^2(V_{J_\star(\lambda)},V_{J_\star^c(\lambda)})\geq\BIAS^\FSD(\lambda)\geq\BIAS^\FSD(\sqrt{\lambda^\star})\\ &\geq\Bias_{J_\star^c(\sqrt{\lambda^\star})}= \frac{\sqrt{\lambda^\star}|J_\star^c(\sqrt{\lambda^\star})|}{N+\sqrt{\lambda^\star}p} \geq \frac{1}{4}\min\Bigl\{\frac{\sqrt{\lambda^\star}p}{N},1\Bigr\}.
            \end{align*}Since $\lambda^\star=\frac{N^{1/\alpha}}{p}$, we have $\frac{\sqrt{\lambda^\star}p}{N}= N^{1/\alpha-1}\sqrt{\frac{p}{N^{1/\alpha}}}\geq N^{1/\alpha-1}\log(\frac{p}{N^{1/\alpha}})$. Moreover, $\log N\leq\frac{2\alpha}{e(\alpha-1)}N^{(\alpha-1)/(2\alpha)}$ and $p\leq N^\iota$, so the choice of $N_0$ gives $N^{1/\alpha-1}\log(\frac{p}{N^{1/\alpha}})\leq1$. Thus,
            \begin{align*}
                \forall\lambda\geq\sqrt{\lambda^\star},\qquad &r^2(V_{J_\star(\lambda)},V_{J_\star^c(\lambda)})\gtrsim_\alpha N^{1/\alpha-1}\log\Bigl(\frac{p}{N^{1/\alpha}}\Bigr).
            \end{align*}
        \end{enumerate}
    \end{enumerate}
    Thus, for every $N>N_0$,
    \begin{align*}
        r^2(V_{J_\star(\lambda^\star)},V_{J_\star(\lambda^\star)^c})\asymp_{\alpha,\iota}\inf\bigl\{r^2(V_{J_\star(\lambda)},V_{J_\star(\lambda)^c}):\lambda>0\bigr\}.
    \end{align*}
\end{proof}

\section{Proof of other propositions}
\subsection{Proof of Proposition~\texorpdfstring{\ref{prop:monotonicity}}{1}}\label{sec:proof_monotonicity}
We fix $N,p\in\mathbb{N}_+$ and $\bbeta^\star\in\Delta_p$, and discuss the effect of varying $\lambda\in(0,\infty)$ on $J_\star(\lambda)$, whose definition we recall:
\begin{align}\tag{\ref{eq:decomposition}}
    \begin{aligned}
        &J_\star(\lambda)=\bigg\{1\leq j\leq p:\beta_j^\star>\frac\lambda{4(N+\lambda p)}\bigg\}.
    \end{aligned}
\end{align}From the discussion in Section~\ref{sec:geometric_perspective}, we know that the indices in $J_\star(\lambda)$ correspond to the coordinates $\beta_j^\star$ estimated by $\hat\bbeta$. Therefore, by definition, for any $k\in[p]$, exactly one of the following three cases holds:
\begin{enumerate}
    \item if $\beta_k^\star=0$, then $k\notin J_\star(\lambda)$ for any $\lambda>0$;
    \item if $\beta_k^\star\geq\frac{1}{4p}$, then $k\in J_\star(\lambda)$ for any $\lambda>0$; and
    \item if $0<\beta_k^\star<\frac{1}{4p}$, then for any $\lambda>0$, $k\in J_\star(\lambda)$ if and only if $\lambda<\frac{4N\beta_k^\star}{1-4p\beta_k^\star}$.
\end{enumerate}
Namely: for any $\lambda>0$, the Add-$\lambda$ estimator $\hat\bbeta^\lambda$ can always assign the coordinates outside the support to the free subspace, and can always assign the sufficiently important coordinates, namely those exceeding $\frac{1}{4p}$, to the estimation subspace.
\begin{proof}[Proof of Proposition~\ref{prop:monotonicity}]
It is easy to see that $\lambda\in(0,\infty)\mapsto J_\star(\lambda)$ is piecewise constant, and that for any $0<\lambda_1\leq \lambda_2$, we have $J_\star(\lambda_2)\subseteq J_\star(\lambda_1)$. We first work on open intervals on which $J_\star(\lambda)$ is constant, splitting them further at $\lambda=1$. First we note that $\lambda\mapsto J_\star(\lambda)$ is right-continuous.
\begin{enumerate}
    \item Within each open interval, we examine the following derivative. For any $j\in J_\star(\lambda)$,
    \begin{align*}
        &\frac{\mathrm{d}}{\mathrm{d}\lambda}\biggl( \beta_j^\star\log\frac{(N+\lambda p)\beta_j^\star}{N\beta_j^\star+\lambda} + \frac{N\beta_j^\star+\lambda}{N+\lambda p} - \beta_j^\star \biggr) = \beta_j^\star\biggl( \frac{p}{N+\lambda p} - \frac{1}{N\beta_j^\star+\lambda} \biggr) + \frac{N(1-p\beta_j^\star)}{(N+\lambda p)^2}\\
        & = \beta_j^\star\frac{N(p\beta_j^\star-1)}{(N+\lambda p)(N\beta_j^\star+\lambda)} + \frac{N(1-p\beta_j^\star)}{(N+\lambda p)^2} = \frac{N(1-p\beta_j^\star)}{(N+\lambda p)^2}\biggl(1-\frac{\beta_j^\star(N+\lambda p)}{N\beta_j^\star+\lambda}\biggr) = \frac{N\lambda(1-p\beta_j^\star)^2}{(N+\lambda p)^2(N\beta_j^\star+\lambda)}\geq 0,
    \end{align*}and
    \begin{align*}
        \frac{\mathrm{d}}{\mathrm{d}\lambda}\Bias_{J_\star^c(\lambda)} = \frac{N|J_\star^c(\lambda)|}{(N+\lambda p)^2}\geq 0.
    \end{align*}Therefore,
    \begin{align*}
        \frac{\mathrm{d}}{\mathrm{d}\lambda}\BIAS^\FSD(\lambda) = \sum_{j\in J_\star(\lambda)}\frac{\mathrm{d}}{\mathrm{d}\lambda}\biggl[\beta_j^\star\log\frac{(N+\lambda p)\beta_j^\star}{N\beta_j^\star+\lambda}+\frac{N\beta_j^\star+\lambda}{N+\lambda p}-\beta_j^\star\biggr]+ \frac{\mathrm{d}}{\mathrm{d}\lambda}\Bias_{J_\star^c(\lambda)}\geq0,
    \end{align*}indicating that on each such interval, $\lambda\mapsto \BIAS^\FSD(\lambda)$ is nondecreasing.

    Similarly,
    \begin{align*}
        \frac{\mathrm{d}}{\mathrm{d}\lambda}\frac{N(\beta_j^\star)^2(1-\beta_j^\star)}{(N\beta_j^\star+\lambda)(N\beta_j^\star+\lambda+1)} = -\frac{N(\beta_j^\star)^2(1-\beta_j^\star)(2N\beta_j^\star+2\lambda+1)}{(N\beta_j^\star+\lambda)^2(N\beta_j^\star+\lambda+1)^2}\leq 0,
    \end{align*}and
    \begin{align*}
        \frac{\mathrm{d}}{\mathrm{d}\lambda}\biggl(\beta_j^\star(1-\beta_j^\star)^N\log\Bigl(1+\frac{N\beta_j^\star}{\lambda}\Bigr)\biggr) = \beta_j^\star(1-\beta_j^\star)^N\biggl(\frac{1}{\lambda+N\beta_j^\star} - \frac{1}{\lambda}\biggr) = -\frac{N(\beta_j^\star)^2(1-\beta_j^\star)^N}{\lambda(\lambda+N\beta_j^\star)}\leq 0.
    \end{align*}Therefore, on each such interval, $\lambda\mapsto \VAR^\FSD(\lambda)$ is nonincreasing.

    \item Next we only need to determine, at each jump point of $\lambda\in(0,\infty)\mapsto J_\star(\lambda)$, whether the left and right limits at each jump discontinuity of $\BIAS^\FSD(\lambda)$ and $\VAR^\FSD(\lambda)$ satisfy the corresponding monotonicity. To this end, we recall that our discussion in the beginning of this subsection indicates that the jump points of $\lambda\in(0,\infty)\mapsto J_\star(\lambda)$ are
    \begin{align*}
        \mathrm{Jump}:= \mathrm{Jump}(\bbeta^\star) := \biggl\{ \frac{4N\beta_k^\star}{1-4p\beta_k^\star}: k\in[p],\quad 0<\beta_k^\star<\frac{1}{4p} \biggr\}.
    \end{align*}Let $m=|\mathrm{Jump}|$, and let $\mathrm{Jump}=\{\lambda_1,\cdots,\lambda_m\}$ with $\lambda_1<\cdots<\lambda_m$.  For any $1\leq j\leq m$, let
    \begin{align*}
        \BIAS^\FSD(\lambda_j^-) = \lim_{\lambda\to\lambda_j^-}\BIAS^\FSD(\lambda),\quad\mbox{ and }\quad \BIAS^\FSD(\lambda_j^+) = \lim_{\lambda\to\lambda_j^+}\BIAS^\FSD(\lambda),
    \end{align*}Namely, we define $\BIAS^\FSD(\lambda_j^-)$ and $\BIAS^\FSD(\lambda_j^+)$ as the limits of $\BIAS^\FSD(\lambda)$ as $\lambda$ approaches $\lambda_j$ from the left and from the right, respectively. By computation,
    \begin{align*}
        &\BIAS^\FSD(\lambda_j^-) = \sum_{\beta_k^\star\geq\frac{\lambda_j}{4(N+\lambda_j p)}} \biggl( \beta_k^\star \log\frac{(N+\lambda_jp)\beta_k^\star}{N\beta_k^\star+\lambda_j} + \frac{N\beta_k^\star+\lambda_j}{N+\lambda_jp} - \beta_k^\star \biggr) + \sum_{\beta_k^\star<\frac{\lambda_j}{4(N+\lambda_jp)}}\frac{\lambda_j}{N+\lambda_jp},
    \end{align*}and
    \begin{align*}
        &\BIAS^\FSD(\lambda_j^+) = \sum_{\beta_k^\star>\frac{\lambda_j}{4(N+\lambda_j p)}} \biggl( \beta_k^\star \log\frac{(N+\lambda_jp)\beta_k^\star}{N\beta_k^\star+\lambda_j} + \frac{N\beta_k^\star+\lambda_j}{N+\lambda_jp} - \beta_k^\star \biggr) + \sum_{\beta_k^\star\leq \frac{\lambda_j}{4(N+\lambda_jp)}}\frac{\lambda_j}{N+\lambda_jp}.
    \end{align*}Therefore,
    \begin{align*}
        &\BIAS^\FSD(\lambda_j^-) - \BIAS^\FSD(\lambda_j^+) = \sum_{\beta_k^\star = \frac{\lambda_j}{4(N+\lambda_j p)}}\Biggl( \beta_k^\star \log\frac{(N+\lambda_jp)\beta_k^\star}{N\beta_k^\star+\lambda_j} + \frac{N\beta_k^\star+\lambda_j}{N+\lambda_jp} - \beta_k^\star - \frac{\lambda_j}{N+\lambda_jp}\Biggr).
    \end{align*}By the definition of $\lambda_j$, $\frac{\lambda_j}{N+\lambda_jp}=4\beta_k^\star$, hence $N+\lambda_j p=\frac{N(1-4p\beta_k^\star)+4Np\beta_k^\star}{1-4p\beta_k^\star} = \frac{N}{1-4p\beta_k^\star}$, and hence
    \begin{align*}
        \frac{N\beta_k^\star+\lambda_j}{N+\lambda_jp} = \frac{(N+\lambda_jp)\beta_k^\star+\lambda_j(1-p\beta_k^\star)}{N+\lambda_jp} = \beta_k^\star + \frac{\lambda_j}{N+\lambda_jp}(1-p\beta_k^\star) = \beta_k^\star + 4\beta_k^\star(1-p\beta_k^\star) = (5-4p\beta_k^\star)\beta_k^\star.
    \end{align*}Then
    \begin{align*}
        &\BIAS^\FSD(\lambda_j^-) - \BIAS^\FSD(\lambda_j^+) = \sum_{\beta_k^\star = \frac{\lambda_j}{4(N+\lambda_j p)}}\beta_k^\star \biggl( -\log\Bigl(5-4p\beta_k^\star\Bigr) + \Bigl(5-4p\beta_k^\star\Bigr)-1-4 \biggr)\\
        &= -\sum_{\beta_k^\star = \frac{\lambda_j}{4(N+\lambda_j p)}}\beta_k^\star \Bigl( \log\Bigl(5-4p\beta_k^\star\Bigr) + 4p\beta_k^\star\Bigr)<0,
    \end{align*}where we have used that $0<4p\beta_k^\star<1$ in the last inequality.

    Similarly, we compute $\VAR^\FSD(\lambda_j^-)$ and $\VAR^\FSD(\lambda_j^+)$ when $\lambda_j\neq 1$. One may compute that
    \begin{align*}
        &\VAR^\FSD(\lambda_j^-) - \VAR^\FSD(\lambda_j^+) = \sum_{\beta_k^\star = \frac{\lambda_j}{4(N+\lambda_j p)}}\biggl(\frac{N(\beta_k^\star)^2(1-\beta_k^\star)}{(N\beta_k^\star+\lambda_j)(N\beta_k^\star+\lambda_j+1)} +\one_{\{\lambda_j<1\}}\beta_k^\star(1-\beta_k^\star)^N\log\Bigl(1+\frac{N\beta_k^\star}{\lambda_j}\Bigr) \biggr),
    \end{align*}which is non-negative. We also check $\lambda=1$, whether or not $1\in\mathrm{Jump}$. We have 
    \begin{align*}
        &\VAR^\FSD(1^-) - \VAR^\FSD(1^+) = \sum_{\beta_k^\star = \frac{1}{4(N+p)}}\biggl(\frac{N(\beta_k^\star)^2(1-\beta_k^\star)}{(N\beta_k^\star+1)(N\beta_k^\star+2)}\biggr) + \sum_{\beta_k^\star \geq \frac{1}{4(N+p)}}\biggl(\beta_k^\star(1-\beta_k^\star)^N\log\Bigl( 1+N\beta_k^\star \Bigr) \biggr),
    \end{align*}which is non-negative.
\end{enumerate}Both functions take their right limits at the switching points, so these calculations prove the claimed monotonicity of $\lambda\in(0,\infty)\mapsto \BIAS^\FSD(\lambda)$ and $\lambda\in(0,\infty)\mapsto\VAR^\FSD(\lambda)$.
\end{proof}

\subsection{Proof of Proposition~\texorpdfstring{\ref{prop:no_phase_transition}}{no phase transition}}\label{sec:proof_prop_no_phase_transition}
\begin{proof}[Proof of Proposition~\ref{prop:no_phase_transition}]
We compare the corresponding $\BIAS^\FSD(\lambda),\VAR^\FSD(\lambda)$ with $\BIAS^\FSD(1),\VAR^\FSD(1)$. Recall that
\begin{align}\tag{\ref{eq:fsd-components}}
    \begin{aligned}&\mathrm{Bias}_{J_\star}=\sum_{j\in J_\star}\biggl[\beta_j^\star\log\frac{(N+\lambda p)\beta_j^\star}{N\beta_j^\star+\lambda}+\frac{N\beta_j^\star+\lambda}{N+\lambda p}-\beta_j^\star\biggr],\quad \mathrm{Bias}_{J_\star^c}=\frac{\lambda|J_\star^c|}{N+\lambda p},\quad\mbox{ and }\\ &\\ &\mathrm{VAR}^{\mathrm{FSD}}=\sum_{j\in J_\star}\biggl[\frac{N(\beta_j^\star)^2(1-\beta_j^\star)}{(N\beta_j^\star+\lambda)(N\beta_j^\star+\lambda+1)}+\mathbf1_{\{\lambda<1\}}\beta_j^\star(1-\beta_j^\star)^N\log\biggl(1+\frac{N\beta_j^\star}{\lambda}\biggr)\biggr].\end{aligned}
\end{align}In particular, we let
\begin{align*}
    \Bias_j(\lambda) =\begin{cases}
        \beta_j^\star\log\frac{(N+\lambda p)\beta_j^\star}{N\beta_j^\star+\lambda}+\frac{N\beta_j^\star+\lambda}{N+\lambda p}-\beta_j^\star,&\mbox{ for }j\in J_\star(\lambda)\\
        \frac{\lambda}{N+\lambda p},&\mbox{otherwise}.
    \end{cases} 
\end{align*}Then $r^2(V_{J_\star(\lambda)},V_{J_\star^c(\lambda)})=\sum_{j=1}^p\Bigl(\Bias_j(\lambda) + \one_{\{j\in J_\star(\lambda)\}}\Var_j(\lambda)\Bigr)$, where $\Var_j(\lambda)$ is defined in Lemma~\ref{lem:jensen-upper}. It suffices to prove
\begin{align}\label{eq:objective_Bias_j_Var_j}
    \forall \frac{1}{2}\leq\lambda\leq 1,\quad \forall j\in[p],\qquad \Bias_j(\lambda)+\one_{\{j\in J_\star(\lambda)\}}\Var_j(\lambda)\asymp \Bias_j(1)+\one_{\{j\in J_\star(1)\}}\Var_j(1)
\end{align}
summing over $j$ then yields the desired conclusion. 

\textbf{Preliminary.} For sets $A,B$, we let $A\triangle B$ be $(A\cap B^c)\cup(A^c\cap B)$. We discuss the following three cases:
\begin{enumerate}
    \item $j\in J_\star(\lambda)\cap J_\star(1)$;
    \item $j\in J_\star^c(\lambda)\cap J_\star(1)$; and
    \item $j\in J_\star(\lambda)\triangle J_\star(1)$.
\end{enumerate}
One may verify the following three facts: for any $\frac{1}{2}\leq\lambda\leq 2$ and any $j\in[p]$, there hold
\begin{equation}\label{eq:no-phase-ratios}
    \begin{aligned}
     \min\{1,\lambda\}  \leq\frac{\lambda(N+p)}{N+\lambda p}\leq\max\{1,\lambda\},&\quad \frac12\leq\frac{(N\beta_j^\star+\lambda)(N+p)}{(N+\lambda p)(N\beta_j^\star+1)}\leq2,\quad\mbox{ and }\quad \frac13\leq\frac{(N\beta_j^\star+1)(N\beta_j^\star+2)}{(N\beta_j^\star+\lambda)(N\beta_j^\star+\lambda+1)}\leq\frac83,\end{aligned}
\end{equation}The last of these facts follows from the monotonic increasing property of $\lambda\mapsto (N\beta_j^\star+\lambda)(N\beta_j^\star+\lambda+1)$, by substituting $\lambda=\frac12$ and $\lambda=2$. By \eqref{eq:no-phase-ratios}, we obtain that
\begin{align}\label{eq:log_negligible}
    \begin{aligned}
        \forall \frac{1}{2}\leq\lambda<1,\quad \forall0<\beta_j^\star<1,\qquad &\frac{\beta_j^\star(1-\beta_j^\star)^N\log\Bigl(1+\frac{N\beta_j^\star}{\lambda}\Bigr)}{\frac{N(\beta_j^\star)^2(1-\beta_j^\star)}{(N\beta_j^\star+\lambda)(N\beta_j^\star+\lambda+1)}}\leq \frac{(N\beta_j^\star+\lambda)(N\beta_j^\star+\lambda+1)}{\lambda}(1-\beta_j^\star)^{N-1}\\
        &\leq 2(N\beta_j^\star+1)^2\exp(1-N\beta_j^\star)\leq 8,
    \end{aligned}
\end{align}where we have used $\log(1+x)\leq x$, $\frac{1}{\lambda}\leq 2$, $\lambda+1\leq 2$, and $(1-\beta_j^\star)^{N-1}\leq \exp(1-N\beta_j^\star)$. When $\beta_j^\star\in\{0,1\}$, $\Var_j(\lambda)=0$ for any $j\in[p]$. When $\lambda\geq 1$, $\mathbf1_{\{\lambda<1\}}\beta_j^\star(1-\beta_j^\star)^N\log\Bigl(1+\frac{N\beta_j^\star}{\lambda}\Bigr)=0$. Therefore, we summarize from \eqref{eq:log_negligible} that
\begin{align}\label{eq:equivalence_Var_j}
    \forall \frac{1}{2}\leq\lambda\leq 2,\quad \forall j\in[p],\qquad \Var_j(\lambda)\asymp \Var_j(1).
\end{align}
Moreover, we recall that
\begin{align}\tag{\ref{eq:identity_Bias_j}}
    \forall \lambda>0,\quad \forall j\in J_\star(\lambda),\qquad&\beta_j^\star\log\frac{(N+\lambda p)\beta_j^\star}{N\beta_j^\star+\lambda}+\frac{N\beta_j^\star+\lambda}{N+\lambda p}-\beta_j^\star = \frac{\lambda^2(1-p\beta_j^\star)^2}{(N+\lambda p)^2}\int_0^1\frac{t\mathrm{d}t}{(1-t)\beta_j^\star + t\frac{N\beta_j^\star+\lambda}{N+\lambda p}}.
\end{align} 
The last piece is the following equivalence
\begin{align}\label{eq:equivalence_bias_j}
    \forall \frac{\lambda}{4(N+\lambda p)} < \beta_j^\star\leq \frac{\lambda}{2(N+\lambda p)},\qquad \Bias_j(\lambda)\asymp \frac{\lambda}{N+\lambda p},\mbox{ and } \frac{N(\beta_j^\star)^2(1-\beta_j^\star)}{(N\beta_j^\star+\lambda)(N\beta_j^\star+\lambda+1)} \leq \frac{\lambda}{2(N+\lambda p)}.
\end{align}Indeed, for any $\beta_j^\star$ such that $\frac{\lambda}{4(N+\lambda p)} < \beta_j^\star\leq \frac{\lambda}{2(N+\lambda p)}$, we have $p\beta_j^\star\leq\frac{1}{2}$, hence
\begin{align*}
    &\frac{\lambda}{2(N+\lambda p)} \leq \frac{N\beta_j^\star+\lambda}{N+\lambda p} - \beta_j^\star = \frac{\lambda(1-p\beta_j^\star)}{N+\lambda p}\leq \frac{\lambda}{N+\lambda p}.
\end{align*}Moreover, since the left-hand-side of the above displayed inequity is non-negative, we obtain $\beta_j^\star\leq\frac{N\beta_j^\star+\lambda}{N+\lambda p}$, hence for any $0\leq t\leq 1$, the convex combination satisfies $(1-t)\beta_j^\star + t\frac{N\beta_j^\star+\lambda}{N+\lambda p}\leq\frac{N\beta_j^\star+\lambda}{N+\lambda p}$. Together with \eqref{eq:identity_Bias_j},
\begin{align*}
    \Bias_j(\lambda) &= \Bigl( \frac{N\beta_j^\star+\lambda}{N+\lambda p} - \beta_j^\star \Bigr)^2\int_0^1\frac{t\mathrm{d}t}{(1-t)\beta_j^\star + t\frac{N\beta_j^\star+\lambda}{N+\lambda p} } \geq \Bigl( \frac{N\beta_j^\star+\lambda}{N+\lambda p} - \beta_j^\star \Bigr)^2\frac{N+\lambda p}{N\beta_j^\star+\lambda}\int_0^1 t\mathrm{d}t \\
    &\geq \frac{\Bigl(\frac{N\beta_j^\star+\lambda}{N+\lambda p}-\beta_j^\star\Bigr)^2}{2\frac{N\beta_j^\star+\lambda}{N+\lambda p}} \geq \frac{1}{4}\biggl( \frac{N\beta_j^\star+\lambda}{N+\lambda p} - \beta_j^\star \biggr) \geq \frac{\lambda}{8(N+\lambda p)}.
\end{align*}For the upper side, we use $\frac{N\beta_j^\star+\lambda}{N+\lambda p}\geq\beta_j^\star$, hence the logarithmic term in $\Bias_j(\lambda)$ is non-positive, and consequently $\Bias_j(\lambda)\leq \frac{\lambda}{N+\lambda p}$.

\textbf{Three cases.}
\begin{enumerate}
    \item When $j\in J_\star(\lambda)\cap J_\star(1)$. We first compare $\Bias_j(\lambda)$. By \eqref{eq:no-phase-ratios}, we have
    \begin{align*}
        \forall\frac{1}{2}\leq\lambda\leq2,\qquad \frac{1}{2}\frac{N\beta_j^\star+1}{N+p}\leq \frac{N\beta_j^\star+\lambda}{N+\lambda p} \leq 2\frac{N\beta_j^\star+1}{N+p},\quad\mbox{ and }\quad \frac{1}{4}\leq\frac{\lambda^2(N+p)^2}{(N+\lambda p)^2}\leq 4.
    \end{align*}Combining with \eqref{eq:identity_Bias_j}, we obtain that $\Bias_j(\lambda)\asymp \Bias_j(1)$ for any $\frac{1}{2}\leq\lambda\leq 2$. The comparison of $\Var_j(\lambda)$ and $\Var_j(1)$ is implied by \eqref{eq:equivalence_Var_j}.

    \item When $j\in J_\star^c(\lambda)\cap J_\star^c(1)$. To prove \eqref{eq:objective_Bias_j_Var_j}, we only need to show that $\Bias_j(\lambda)\asymp\Bias_j(1)$, for $j\in J_\star^c(\lambda)\cap J_\star^c(1)$,  that is, to prove $\frac{\lambda}{N+\lambda p}\asymp \frac{1}{N+p}$. Notice that by \eqref{eq:no-phase-ratios},
    \begin{align*}
        \frac{1}{2}\leq \frac{\Bias_j(\lambda)}{\Bias_j(1)} = \frac{\lambda(N+p)}{N+\lambda p}\leq 2,
    \end{align*}the case is verified.

    \item When $j\in J_\star(\lambda)\triangle J_\star(1)$. We further separate into two cases.
    \begin{enumerate}
        \item When $j\in J_\star(\lambda)\backslash J_\star(1)$. Since $\lambda\mapsto J_\star(\lambda)$ is non-increasing, this implies $\frac{1}{2}\leq\lambda< 1$, and to prove \eqref{eq:objective_Bias_j_Var_j}, we only need to prove
        \begin{align*}
            \Bias_j(1)\asymp \Bias_j(\lambda)+\Var_j(\lambda).
        \end{align*}By $j\in J_\star(\lambda)\backslash J_\star(1)$, we know that $\frac{\lambda}{4(N+\lambda p)}<\beta_j^\star\leq\frac{1}{4(N+p)}$. Since $\frac{1}{2}\leq\lambda<1$, we have $\frac{1}{2}\leq \frac{\lambda(N+p)}{N+\lambda p}\leq 1$, hence  \eqref{eq:equivalence_bias_j} implies
        \begin{align}\tag{\ref{eq:equivalence_bias_j}}
            \Bias_j(\lambda)\asymp \frac{\lambda}{N+\lambda p},\mbox{ and } \frac{N(\beta_j^\star)^2(1-\beta_j^\star)}{(N\beta_j^\star+\lambda)(N\beta_j^\star+\lambda+1)} \leq \frac{\lambda}{2(N+\lambda p)}.
        \end{align}Together with \eqref{eq:log_negligible}, we have
        \begin{align*}
            &\Var_j(\lambda) = \frac{N(\beta_j^\star)^2(1-\beta_j^\star)}{(N\beta_j^\star+\lambda)(N\beta_j^\star+\lambda+1)}+\mathbf1_{\{\lambda<1\}}\beta_j^\star(1-\beta_j^\star)^N\log\bigg(1+\frac{N\beta_j^\star}{\lambda}\bigg)\lesssim \frac{\lambda}{2(N+\lambda p)}.
        \end{align*}Therefore,
        \begin{align*}
            \frac{\lambda}{8(N+\lambda p)}\lesssim \Bias_j(\lambda) + \Var_j(\lambda)\leq\lesssim\frac{6}{N+\lambda p}.
        \end{align*}
        \item When $j\in J_\star(1)\backslash J_\star(\lambda)$. This case is similar to the former case, thus omitted.
    \end{enumerate}
\end{enumerate}
\end{proof}

\section{Auxiliary lemmas}\label{sec:auxiliary}

Lemma~\ref{lem:moment_N_j}, Lemma~\ref{lem:integral_log}, Lemma~\ref{lem:identity}, Lemma~\ref{lem:upper_expectation} and Lemma~\ref{lem:covariance} below are used to prove Lemma~\ref{lem:jensen-upper}.

\begin{Lemma}\label{lem:moment_N_j}
    For any $N,p\in\bN_+$, any $\bbeta^\star\in\Delta_p$, any $1\leq j\leq p$, and any $s>0$,
    \begin{align*}
        \bE[(N_j-N\beta_j^\star)^2] = N\beta_j^\star(1-\beta_j^\star),\quad\mbox{ and }\quad \bE[(N_j-N\beta_j^\star)^2(N_j+s)]=N\beta_j^\star(1-\beta_j^\star)(N\beta_j^\star+s+1-2\beta_j^\star).
    \end{align*}
\end{Lemma}
\begin{proof}[Proof of Lemma~\ref{lem:moment_N_j}]
     The distribution of $N_j$ is $\mathrm{Binomial}(N,\beta_j^\star)$, hence $\bE(N_j-N\beta_j^\star)=0$. For $0<\beta_j^\star<1$, $\bE[(N_j-N\beta_j^\star)^2] = \sum_{i=1}^N\bE[(\one_{\{X_i=j\}}-\beta_j^\star)^2] = N\beta_j^\star(1-\beta_j^\star)$, and similarly, $\bE[(N_j-N\beta_j^\star)^3] = N\beta_j^\star(1-\beta_j^\star)(1-2\beta_j^\star)$. Therefore, by the identity $N_j+s=(N_j-N\beta_j^\star)+(N\beta_j^\star+s)$, there holds $(N_j-N\beta_j^\star)^2(N_j+s) = (N_j-N\beta_j^\star)^3 + (N\beta_j^\star+s)(N_j-N\beta_j^\star)^2$, and consequently,
\begin{align*}
    \forall s>0,\qquad \bE[(N_j-N\beta_j^\star)^2(N_j+s)] &= N\beta_j^\star(1-\beta_j^\star)(1-2\beta_j^\star) + (N\beta_j^\star+s)N\beta_j^\star(1-\beta_j^\star)\\
    &=N\beta_j^\star(1-\beta_j^\star)(N\beta_j^\star+s+1-2\beta_j^\star).
\end{align*}
\end{proof}

\begin{Lemma}\label{lem:integral_log}
    For any $N,p\in\bN_+$, any $1\leq j\leq p$, any $\lambda>0$ and any $\bbeta^\star\in\Delta_p$,
    \begin{align}\label{eq:resolvent}
    \log(N\beta_j^\star+\lambda)-\bE\log(N_j+\lambda)=\int_\lambda^\infty\frac{\bE\big[\frac{(N_j-N\beta_j^\star)^2}{N_j+s}\big]}{(N\beta_j^\star+s)^2}\,\mathrm ds.
    \end{align}
\end{Lemma}
\begin{proof}
    On the interval $\{s\in\mathbb{R}:s\geq\lambda\}$, using the fundamental theorem of calculus, we obtain
\begin{align*}
    \log\bigl(N\beta_j^\star+\lambda\bigr) - \log\bigl(N_j+\lambda\bigr) = &\int_\lambda^\infty\Bigl(\frac{1}{N_j+s}-\frac{1}{N\beta_j^\star+s}\Bigr)\mathrm{d}s.
\end{align*}Taking expectations on both sides and interchanging the integral and the expectation, we only need to examine $\mathbb{E}\bigl[\frac{1}{N_j+s}-\frac{1}{N\beta_j^\star+s}\bigr]$. Note the following identity
\begin{align*}
    \frac{1}{N_j+s}-\frac{1}{N\beta_j^\star+s} + \frac{N_j-N\beta_j^\star}{(N\beta_j^\star+s)^2} = \frac{(N_j-N\beta_j^\star)^2}{(N_j+s)(N\beta_j^\star+s)^2}.
\end{align*}
Taking expectations on both sides together with the fact that $\bE[N_j]=N\beta_j^\star$ gives
\begin{align*}
    \bE\Bigl[\frac{(N_j-N\beta_j^\star)^2}{(N_j+s)((N\beta_j^\star+s)^2)}\Bigr] = \bE\Bigl[\frac{1}{N_j+s}\Bigr] - \frac{1}{N\beta_j^\star+s}.
\end{align*}
As a result,
\begin{align}
    \log(N\beta_j^\star+\lambda)-\bE\log(N_j+\lambda)=\int_\lambda^\infty\frac{\bE\big[\frac{(N_j-N\beta_j^\star)^2}{N_j+s}\big]}{(N\beta_j^\star+s)^2}\,\mathrm ds.\tag{\ref{eq:resolvent}}
\end{align}
\end{proof}

\begin{Lemma}\label{lem:identity}
    For any $N,p\in\bN_+$, any $j\in[p]$, any $0\leq k\leq N$, and any $\bbeta^\star\in\Delta_p$
    \begin{align}\label{eq:binomial_identity}
        \begin{aligned}
            &(k-N\beta_j^\star)\binom{N}{k}(\beta_j^\star)^k(1-\beta_j^\star)^{N-k}\\
            &= N\beta_j^\star(1-\beta_j^\star)\Bigl( \binom{N-1}{k-1}(\beta_j^\star)^{k-1}(1-\beta_j^\star)^{N-k} - \binom{N-1}{k}(\beta_j^\star)^k(1-\beta_j^\star)^{N-1-k} \Bigr),
        \end{aligned}
    \end{align}where, when $k=0$ ,$\binom{N-1}{k-1}(\beta_j^\star)^{k-1}(1-\beta_j^\star)^{N-k} = 0$; and when $k=N$, $\binom{N-1}{k}(\beta_j^\star)^k(1-\beta_j^\star)^{N-1-k}=0$.
\end{Lemma}
\begin{proof}[Proof of Lemma~\ref{lem:identity}]
Notice that for any $0\leq k\leq N$, $k-N\beta_j^\star=k(1-\beta_j^\star) - (N-k)\beta_j^\star$, $k\binom{N}{k}=N\binom{N-1}{k-1}$, and $(N-k)\binom{N}{k}=N\binom{N-1}{k}$; hence
    \begin{align}\tag{\ref{eq:binomial_identity}}
        \begin{aligned}
            &(k-N\beta_j^\star)\binom{N}{k}(\beta_j^\star)^k(1-\beta_j^\star)^{N-k}\\
            &= N\beta_j^\star(1-\beta_j^\star)\Bigl( \binom{N-1}{k-1}(\beta_j^\star)^{k-1}(1-\beta_j^\star)^{N-k} - \binom{N-1}{k}(\beta_j^\star)^k(1-\beta_j^\star)^{N-1-k} \Bigr),
        \end{aligned}
    \end{align}
\end{proof}

\begin{Lemma}\label{lem:upper_expectation}
    For any $N,p\in\bN_+$, any $1\leq j\leq p$, and any $\bbeta^\star\in\Delta_p$,
    \begin{align}\label{eq:binomial-reciprocal}
        &\bE\frac{(N_j-N\beta_j^\star)^2}{N_j+1} \leq \frac{3N\beta_j^\star(1-\beta_j^\star)}{N\beta_j^\star+1}.
    \end{align}
\end{Lemma}
\begin{proof}[Proof of Lemma~\ref{lem:upper_expectation}]
    Since $\bP(N_j=k)=\binom{N}{k}(\beta_j^\star)^k(1-\beta_j^\star)^{N-k}$, by Lemma~\ref{lem:identity}, there holds
    \begin{align*}
        \bE\frac{(N_j-N\beta_j^\star)^2}{N_j+1} &= \sum_{k=0}^N\frac{(k-N\beta_j^\star)^2}{k+1}\binom{N}{k}(\beta_j^\star)^k(1-\beta_j^\star)^{N-k} \\
        &= N\beta_j^\star(1-\beta_j^\star)\sum_{k=1}^N\frac{k-N\beta_j^\star}{k+1}\binom{N-1}{k-1}(\beta_j^\star)^{k-1}(1-\beta_j^\star)^{N-k}\\
        &\qquad-N\beta_j^\star(1-\beta_j^\star)\sum_{k=0}^{N-1}\frac{k-N\beta_j^\star}{k+1}\binom{N-1}{k}(\beta_j^\star)^k(1-\beta_j^\star)^{N-1-k}.
    \end{align*}Re-indexing the first summation to $k=0,\cdots,N-1$ gives
    \begin{align*}
    \begin{aligned}
        &\bE\frac{(N_j-N\beta_j^\star)^2}{N_j+1} = N\beta_j^\star(1-\beta_j^\star)\sum_{k=0}^{N-1}\frac{k+1-N\beta_j^\star}{k+2}\binom{N-1}{k}(\beta_j^\star)^k(1-\beta_j^\star)^{N-1-k} \\
        & \qquad -N\beta_j^\star(1-\beta_j^\star)\sum_{k=0}^{N-1}\frac{k-N\beta_j^\star}{k+1}\binom{N-1}{k}(\beta_j^\star)^k(1-\beta_j^\star)^{N-1-k} \\
        &= N\beta_j^\star(1-\beta_j^\star)\sum_{k=0}^{N-1}\binom{N-1}{k}(\beta_j^\star)^k(1-\beta_j^\star)^{N-1-k}\Bigl(\frac{k+1-N\beta_j^\star}{k+2}-\frac{k-N\beta_j^\star}{k+1}\Bigr) \\
        &=N\beta_j^\star(1-\beta_j^\star)(N\beta_j^\star+1)\sum_{k=0}^{N-1}\frac{\binom{N-1}{k}(\beta_j^\star)^k(1-\beta_j^\star)^{N-1-k}}{(k+1)(k+2)}.
    \end{aligned}
    \end{align*}Now since $\frac{1}{(k+1)(k+2)}=\int_0^1t^k(1-t)\mathrm{d}t$,
    \begin{align*}
        &\sum_{k=0}^{N-1}\frac{\binom{N-1}{k}(\beta_j^\star)^k(1-\beta_j^\star)^{N-1-k}}{(k+1)(k+2)} = \int_0^1(1-t)\sum_{k=0}^{N-1}\binom{N-1}{k}(\beta_j^\star t)^k(1-\beta_j^\star)^{N-1-k}\mathrm{d}t\\
        &=\int_0^1(1-t)\bigl(1-\beta_j^\star+\beta_j^\star t\bigr)^{N-1}\mathrm{d}t = \Bigl( \frac{(1-\beta_j^\star+\beta_j^\star t)^N}{N(\beta_j^\star)^2} - \frac{(1-\beta_j^\star+\beta_j^\star t)^{N+1}}{(N+1)(\beta_j^\star)^2} \Bigr)\bigg|_{t=0}^{t=1} = \frac{1-(1-\beta_j^\star)^N(1+N\beta_j^\star)}{N(N+1)(\beta_j^\star)^2}.
    \end{align*}Substituting this back gives
    \begin{align*}
        &\bE\frac{(N_j-N\beta_j^\star)^2}{N_j+1} = (1-\beta_j^\star)(N\beta_j^\star+1)\frac{1-(1-\beta_j^\star)^N(1+N\beta_j^\star)}{(N+1)(\beta_j^\star)}.
    \end{align*}The elementary inequality gives $1-(1-\beta_j^\star)^N(1+N\beta_j^\star) \leq \min\{1,\frac{1}{2}N(N+1)(\beta_j^\star)^2\}$, hence
    \begin{align*}
        &\bE\frac{(N_j-N\beta_j^\star)^2}{N_j+1} \leq \frac{(1-\beta_j^\star)(N\beta_j^\star+1)}{(N+1)\beta_j^\star}\min\Bigl\{1,\frac{1}{2}N(N+1)(\beta_j^\star)^2\Bigr\} = N\beta_j^\star(1-\beta_j^\star)(N\beta_j^\star+1)\min\Bigl\{\frac{1}{2},\frac{1}{N(N+1)(\beta_j^\star)^2}\Bigr\}\\
        &\leq N\beta_j^\star(1-\beta_j^\star)(N\beta_j^\star+1)\min\Bigl\{\frac{1}{2},\frac{1}{(N\beta_j^\star)^2}\Bigr\}.
    \end{align*}When $N\beta_j^\star\leq\sqrt{2}$, $\frac{1}{2}\leq\frac{3}{(N\beta_j^\star+1)^2}$; when $N\beta_j^\star>\sqrt{2}$, $\frac{1}{(N\beta_j^\star)^2} = \frac{(1+1/(N\beta_j^\star))^2}{(N\beta_j^\star+1)^2}\leq \frac{3}{(N\beta_j^\star+1)^2}$. Therefore,
    \begin{align}\tag{\ref{eq:binomial-reciprocal}}
        &\bE\frac{(N_j-N\beta_j^\star)^2}{N_j+1} \leq \frac{3N\beta_j^\star(1-\beta_j^\star)}{N\beta_j^\star+1}.
    \end{align}
\end{proof}

\begin{Lemma}\label{lem:covariance}
    For every $j\in[p]$ and $\lambda\geq 1$,
    \begin{align*}
        &\Cov(N_j,\log(N_j+\lambda)) \leq \frac{2N\beta_j^\star(1-\beta_j^\star)}{N\beta_j^\star+\lambda}.
    \end{align*}
\end{Lemma}
\begin{proof}[Proof of Lemma~\ref{lem:covariance}]
    By $\bP(N_j=k)=\binom{N}{k}(\beta_j^\star)^k(1-\beta_j^\star)^{N-k}$ again,
    \begin{align*}
        &\Cov(N_j,\log(N_j+\lambda)) = \bE\bigl[ (N_j-N\beta_j^\star)\log(N_j+\lambda) \bigr] = \sum_{k=0}^N(k-N\beta_j^\star)\binom{N}{k}(\beta_j^\star)^k(1-\beta_j^\star)^{N-k}\log(k+\lambda).
    \end{align*}By \eqref{eq:binomial_identity}, together with re-indexing,
    \begin{align*}
        \Cov(N_j,\log(N_j+\lambda)) &= N\beta_j^\star(1-\beta_j^\star)\sum_{k=1}^N\binom{N-1}{k-1}(\beta_j^\star)^{k-1}(1-\beta_j^\star)^{N-k}\log(k+\lambda) \\
        &\qquad- N\beta_j^\star(1-\beta_j^\star)\sum_{k=0}^{N-1}\binom{N-1}{k}(\beta_j^\star)^k(1-\beta_j^\star)^{N-1-k}\log(k+\lambda)\\
        &=N\beta_j^\star(1-\beta_j^\star)\sum_{k=0}^{N-1}\binom{N-1}{k}(\beta_j^\star)^k(1-\beta_j^\star)^{N-1-k}\Bigl(\log(k+1+\lambda)-\log(k+\lambda)\Bigr)\\
        &=N\beta_j^\star(1-\beta_j^\star)\sum_{k=0}^{N-1}\binom{N-1}{k}(\beta_j^\star)^k(1-\beta_j^\star)^{N-1-k}\log\Bigl(1+\frac{1}{k+\lambda}\Bigr).
    \end{align*}Since $\lambda\geq 1$, there holds $\log(1+\frac{1}{k+\lambda})\leq \frac{1}{k+\lambda}\leq\min\{\frac{1}{\lambda},\frac{1}{k+1}\}$. We first have $\sum_{k=0}^{N-1}\binom{N-1}{k}(\beta_j^\star)^k(1-\beta_j^\star)^{N-1-k}\log\bigl(1+\frac{1}{k+\lambda}\bigr)\leq \frac{1}{\lambda}$, since the binomial sum equals $1$. On the other hand, by $\frac{1}{k+1}=\int_0^1t^k\mathrm{d}t$,
        \begin{align*}
            &\sum_{k=0}^{N-1}\binom{N-1}{k}\frac{(\beta_j^\star)^k(1-\beta_j^\star)^{N-1-k}}{k+1} = \int_0^1\sum_{k=0}^{N-1}\binom{N-1}{k}(\beta_j^\star t)^k(1-\beta_j^\star)^{N-1-k}\mathrm{d}t =\int_0^1(1-\beta_j^\star+\beta_j^\star t)^{N-1}\mathrm{d}t\\
            &=\frac{1-(1-\beta_j^\star)^N}{N\beta_j^\star}\leq \frac{1}{N\beta_j^\star}.
        \end{align*}Combining these bounds,
        \begin{align*}
            &\Cov(N_j,\log(N_j+\lambda)) \leq N\beta_j^\star(1-\beta_j^\star)\min\Bigl\{\frac{1}{\lambda},\frac{1}{N\beta_j^\star}\Bigr\} = \frac{N\beta_j^\star(1-\beta_j^\star)}{\max\{\lambda,N\beta_j^\star\}} \leq \frac{2N\beta_j^\star(1-\beta_j^\star)}{N\beta_j^\star+\lambda}.
        \end{align*}
\end{proof}

The following lemma is used for the geometric perspective on the estimation error of the Add-$\lambda$ estimator in Section~\ref{sec:geometric_perspective}. These facts can be found in standard references such as \cite{amari_information_2016,polyanskiywu}. We include a proof for completeness.
\begin{Lemma}\label{lem:angle}
    Let $p\geq 2$. For any $\bbeta,\bbeta'\in\Delta_p$, let $H(\bbeta,\bbeta'):=\bigl(\frac{1}{2}\sum_{j=1}^p(\sqrt{\beta_j}-\sqrt{\beta_j'})^2\bigr)^{1/2}$ denote the Hellinger distance between $\bbeta$ and $\bbeta'$. Then the following holds.
    \begin{enumerate}
        \item $\cos\angle(2\sqrt{\bbeta},2\sqrt{\bbeta'})=1-H^2(\bbeta,\bbeta')$;
        \item $\KL(\bbeta\|\bbeta')\geq 2H^2(\bbeta,\bbeta')$; and 
        \item for any fixed $\bbeta'\in\Delta_p$ with $\beta_j'>0$ for $1\leq j\leq p$, $\lim_{\bbeta\to\bbeta'}\KL(\bbeta\|\bbeta')/(\angle(2\sqrt{\bbeta},2\sqrt{\bbeta'}))^2=2$.
    \end{enumerate}
\end{Lemma}
\begin{proof}[Proof of Lemma~\ref{lem:angle}]
    Since $2\sqrt{\bbeta},2\sqrt{\bbeta'}\in S_2^{p-1}(\vzero;2)$,
    \begin{align*}
        \cos\angle(2\sqrt{\bbeta},2\sqrt{\bbeta'}) = \frac{1}{4}\langle2\sqrt{\bbeta},2\sqrt{\bbeta'}\rangle = \sum_{j=1}^p\sqrt{\beta_j\beta_j'} = 1-\frac{1}{2}\sum_{j=1}^p (\sqrt{\beta_j}-\sqrt{\beta_j'})^2 = 1-H^2(\bbeta,\bbeta').
    \end{align*}Item~\emph{2.} follows from \cite[Equation 7.33]{polyanskiywu}. Next we show item~\emph{3.}. Since $\beta_j\log\frac{\beta_j}{\beta_j'}=(\beta_j-\beta_j') + \frac{1}{2}\frac{(\beta_j-\beta_j')^2}{\beta_j'} + o((\beta_j-\beta_j')^2)$, we obtain that $\KL(\bbeta\|\bbeta')=(\frac{1}{2}+o(1))\sum_{j=1}^p\frac{(\beta_j-\beta_j')^2}{\beta_j'}$. Similarly, $H^2(\bbeta,\bbeta')=(\frac{1}{8}+o(1))\sum_{j=1}^p\frac{(\beta_j-\beta_j')^2}{\beta_j'}$. Therefore, $\KL(\bbeta\|\bbeta')/H^2(\bbeta,\bbeta')\to 4$ as $\bbeta\to\bbeta'$, Also, $1-\cos t=(\frac12+o(1))t^2$ as $t\to0$, so item~\emph{3} follows from item~\emph{1}.
\end{proof}
\bibliographystyle{alpha}
\bibliography{references}
\end{document}